\documentclass[11pt]{article} 

\usepackage[T1,LGR]{fontenc}
\usepackage[utf8]{inputenc} 
\usepackage[english,greek]{babel} 
\usepackage{amsmath}
\usepackage{amssymb}
\usepackage{txfonts}
\usepackage{mathdots}
\usepackage[classicReIm]{kpfonts}
\usepackage{graphicx}
\usepackage{etoolbox}
    \AfterEndEnvironment{hangparas}{\addvspace{0.67\baselineskip}}
    \usepackage[notquote]{hanging}
\usepackage{textalpha}
\usepackage{alphabeta}
\usepackage{oplotsymbl}
\usepackage[breaklinks=true]{hyperref}
\usepackage{breakcites}
\usepackage{geometry}
\begin{document}

\selectlanguage{english} 


\date{}

\title{\vspace{-3cm}}\maketitle

\noindent \textbf{\Large Real time large scale \textit{in vivo} observations by light-sheet micro-scopy reveal intrinsic synchrony, plasticity and growth cone dynamics of midline crossing axons at the ventral floor plate of the zebrafish spinal cord}

\vspace{\baselineskip}

\noindent S{\o}ren S. L. Andersen${}^{1,*}$

\vspace{\baselineskip}

\noindent \textbf{Abstract}

\noindent Axonal growth and guidance at the ventral floor plate is here followed \textit{in vivo} in real time at high resolution by light-sheet microscopy along several hundred micrometers of the zebrafish spinal cord. The recordings show the strikingly stereotyped spatio-temporal control that governs midline crossing. Commissural axons are observed crossing the ventral floor plate midline perpendicularly at about 20 microns/h, in a manner dependent on the Robo3 receptor and with a growth rate minimum around the midline, confirming previous observations. At guidance points, commissural axons are seen to decrease their growth rate and growth cones increase in size. Commissural filopodia appear to interact with the nascent neural network, and thereby trigger immediate plastic and reversible sinusoidal-shaped bending movements of neighboring commissural shafts. Ipsilateral axons extend concurrently, but straight and without bends, at three to six times higher growth rates than commissurals, indicating they project their path on a substrate-bound surface rather than relying on diffusible guidance cues. Growing axons appeared to be under stretch, an observation that is of relevance for tension-based models of cortical morphogenesis. The \textit{in vivo} observations provide for a discussion of the current distinction between substrate-bound and diffusible guidance cues. The study applies the transparent zebrafish model that provides an experimental model system to explore further the cellular, molecular and physical mechanisms involved during axonal growth, guidance and midline crossing through a combination of \textit{in vivo} and \textit{in vitro} approaches.

\vspace{\baselineskip}
\vspace{\baselineskip}

\begin{small}

\vspace{\baselineskip}

\noindent \textbf{${}^{1}$}The research herein was 2015-2016 carried out by S. Andersen at Uppsala University of Sweden (UU, https://www.UU.se/) at the Swedish Science for Life Laboratory (https://www.SciLifeLab.se/) BioVis Facility. S. Andersen is not an affiliate of UU or SciLifeLab.

\vspace{\baselineskip} 

\noindent \textbf{*}Correspondence: andersen.ssla@gmail.com. ORCID: https://orcid.org/0000-0002-3625-1640. \end{small}

\eject

\noindent \textbf{1. Introduction}

How axons are wired up in an organism is an open and interesting question. In the spinal cord, axons specified as ipsilateral are found on either side of the midline (ipsilateral, contralateral), whereas the commissural axons cross the midline and connect the two body halves (de Ramon Franc\`{a}s et al., 2017; Sakai and Kaprielian, 2012; Graphical Abstract in the accompanying Supporting Information). The midline separates the right and left body halves and represents a crucial guidance point for the growing axons. Errors of midline guidance can have deleterious effects on further development of the organism (Comer et al., 2019; Comer et al., 2015; de Ramon Franc\`{a}s et al., 2017; Tessier-Lavigne and Goodman, 1996). In light hereof, it is not surprising that many of the known axonal guidance receptor-ligand pairs have been characterized in conjunction with studies of axons crossing the midline; including the classical four sets of receptor-ligand molecules: Robos-Slits, DCCs-Netrins, Plexins-Semaphorins and Ephs-Ephrins (Ch\'{e}dotal, 2019; de Ramon Franc\`{a}s et al., 2017; Kolodkin and Tessier-Lavigne, 2011; Short et al., 2021; Stoeckli, 2018; Tessier-Lavigne and Goodman, 1996). 

Technically, neuronal wiring at the cellular scale has until now mostly been studied with various widefield, confocal and two photon microscopy techniques and often using discretely fixed samples. Live high-resolution real time studies of axonal outgrowth \textit{in vivo} have particularly been done on individual axons during pathfinding in the visual system. With its anatomically distinct and well localized axonal pathways, contained in small narrow volumes, confocal and two photon microscopy studies are facilitated (Akin and Zipursky, 2016; Bak and Fraser, 2003; Gaynes et al., 2015; Hutson and Chien, 2002; Mason and Wang, 1997; Nagel et al., 2015; Pittman et al., 2008; Sretavan and Reichardt, 1993). In the spinal cord, individual sensory Rohon-Beard axonal growth has been followed live \textit{in vivo }by widefield and confocal microscopy in the zebrafish (Andersen and Halloran, 2012; Lee et al., 2017; Liu and Halloran, 2005), and individual motoneurons by confocal microscopy in the zebrafish and \textit{Xenopus }(Bremer and Granato, 2016; Plazas et al., 2013) and widefield in \textit{Drosophila }(Murray and Whitington, 1999). \textit{Ex vivo} cultures of chicken spinal cord (Dumoulin et al., 2021), a refinement of the open-book preparation (Bovolenta and Dodd, 1990; Yaginuma et al., 1991), have provided live data on axonal growth and guidance, but live \textit{in vivo} studies of axonal growth and guidance in the chicken model system remain unfeasible with current technology and both genetic and molecular perturbation possibilities are also limited. Though there are compelling reasons to study axon guidance with the \textit{Xenopus} model (reviewed by Erdogan et al., 2021; Erdogan et al., 2016), a live imaging study of \textit{Xenopus} spinal cord floor plate commissurals was based mostly on fixed samples (Moon and Gomez, 2005) because the closely apposed notochord obscured live imaging, just as mouse commissural axon studies also rely on fixed samples (e.g. studies by Brose et al., 1999; Comer et al., 2015; Jaworski et al., 2015).

Direct real time live \textit{in vivo} recordings showing how the many commissural spinal cord axons simultaneously are crossing the midline are missing. Having such information would potentially give new spatio-temporal insights about development as well as the dynamics and mechanisms of axonal wiring on an organismal scale. This study was initiated in an attempt to fill this gap by using the zebrafish model system (Andersen, 2003; Key and Devine, 2003).

The embryonic zebrafish spinal cord is long, curved and transparent, and grows relatively rapidly in all three dimensions during early development. Therefore, to contain physically and image simultaneously several segments of the spinal cord in real time, a large live image acquisition volume is necessary. Recently, the light-sheet technique has rendered such large scale image acquisition possible (Kaufmann et al., 2012; Keller and Ahrens, 2015; Keller et al., 2008; Wan et al., 2019), and one recent real time \textit{in vivo} study has applied a light-sheet microscope, built by the researchers, to investigate angiogenesis in in the zebrafish (Qin et al., 2021), a process that in many ways resembles axon guidance (Fischer et al., 2019). While light-sheet microscopy so far mostly in zebrafish has been used for technically advanced custom-built purposes (e.g.: Wan et al., 2019; Qin et al., 2021; de Vito et al., 2021), with the advent of commercially available light-sheet microscopes and software for live imaging, it has now become doable, straightforward and worth-while to apply this technique to the study of ordinary live \textit{in vivo} cellular biological processes as exemplified in this study. Moreover, data processing and storage is no longer an issue due to available standard software as well as the continuously increasing inexpensive computer storage and processing capacity.

In this light-sheet study, in few seconds, at least hundred micrometers deep areas of around 500 by 500 microns were routinely imaged every 0.2 micron in depth and with a resolution in the range of a few hundred nanometers. Such large high-resolution image volumes are not attainable with confocal or two photon techniques because of the time involved to scan each layer (each z-stack plane) of large volumes. Moreover, the prolonged acquisition times with these techniques, at larger sampling volumes, augment both phototoxicity and may render time lapse imaging prohibitive on a meaningful time scale. Yet, for smaller volumes, confocal and two photon approaches have provided many high-resolution real-time image studies of neuronal dynamics. These techniques have also been used to demonstrate that similar resolution can be reached by light-sheet microscopy (Akin and Zipursky, 2016; Kaufmann et al., 2012; Keller and Ahrens, 2015; Keller et al., 2008; Portera-Cailliau et al., 2005; Weinger et al., 2015).

In the zebrafish spinal cord, outgrowth of axons starts at about 16 hours post fertilization (hpf). In the literature, extensive zebrafish spinal cord nomenclature, electrophysiological studies, and static neuronal mapping of both unidentified, based on morphology and growth patterns, and identified neurons exists (Bernhardt, 1994; Bernhardt et al., 1990; Bernhardt et al., 1992; Drapeau et al., 2002; Hale et al., 2001; Higashijima et al., 2004; Kuwada et al., 1990a; Kuwada et al., 1990b; Lewis and Eisen, 2003). Immediately preceding spinal cord axonal outgrowth, dynamic cell movements during neurulation and neurocoel formation occur and was therefore occasionally also observed in this study due to the developmental spatio-temporal intertwined proximity to commissural outgrowth (Buckley et al., 2013; Cearns et al., 2016; Geldmacher-Voss et al., 2003; Kimmel et al., 1994; Papan and Campos-Ortega, 1994). 

 The model-axon imaged here to study midline crossing \textit{in vivo}, are interneurons expressing GFP from transcription factor promotor dmrt3, previously described in zebrafish as belonging to the dorsal interneuron 6 (dI6) progenitor domain (Satou et al., 2013; Satou et al., 2020; Kishore et al., 2020). This particular type of interneuron has also been studied in mice and horses, indicating a largely uncharacterized both ipsilateral and commissural population (Griener et al., 2017; Staiger et al., 2017).

 Here, a striking spatio-temporal synchrony of commissural axons performing similar simultaneous stereotyped guidance steps in different spinal cord segments is reported. Observing this synchrony required real-time imaging on an organismal scale, and may not be achieved looking at single growing axons by conventional microscopy techniques. The live observations give renewed perspective on the mechanisms of axonal guidance in the spinal cord that provide for a discussion of the current distinction between diffusible long-range versus substrate-bound short-range guidance cues.\eject 

\noindent \textbf{2. Materials and methods}

\vspace{\baselineskip}

\noindent \textbf{2.1 Reagents}

\noindent Zebrafish (RRID: \textit{Danio rerio}) were maintained at 28.5$\mathrm{{}^\circ}$C by the SciLifeLab Zebrafish Facility at UU according to approved institutional methods. Dr. Higashijima, Japan, kindly in 2014 supplied UU with the transgenic Tg(dmrt3:GFP) zebrafish used in this study, expressing GFP under control (GAL4-UAS) of the dmrt3 promotor (Satou et al., 2013; Satou et al., 2020; Kishore et al., 2020 and S. Kishore, C. Satou, K. Ogino, H. Hirata, S.-I. Higashijima, D. Mclean (2015) SfN 421.29/Q1). Embryos were staged according to (Kimmel et al., 1995). Morpholino antisense oligos were translation blocking and produced and designed by Gene Tools{\circledR} (Oregon). 0.25\% trypsin-EDTA (\#25200056), defined trypsin inhibitor (\#R007100) and foetal calf serum (\#10270098) were from Thermofischer Scientific{\circledR}. Tricaine, Phenylthiourea (PTU) and low melt agarose were from Sigma{\circledR}.

\vspace{\baselineskip}

\noindent \textbf{2.2 Summary procedure for recording live from the embryos, larvae and zebrafish}

\noindent For microscopy, the same batch of eggs was used for wild-type and microinjection studies. Eggs were kept at 28.5$\mathrm{{}^\circ}$C; occasionally at 30-32$\mathrm{{}^\circ}$C the first 12 hours to accelerate development (Kimmel et al., 1995). With a stereomicroscope, signs of GFP expression in the spinal cord were routinely monitored from about 12 hpf. About 14-16 hpf, GFP-expressing embryos were mounted in 1\% low-melt agarose dissolved in zebrafish system water with 200 mg/l Tricaine (Kaufmann et al., 2012). The entire embryo, with chorion, was mounted. This was done to give these early embryos the best survival conditions; they are fragile to mechanical handling at this stage and embryos with strong GFP expression in the spinal cord were scarce. The green labelled glass capillaries and plugs for the Carl Zeiss{\circledR} light-sheet microscope were used. Up to four embryos were mounted at one time, on top of each other, and inserted in the light-sheet microscope chamber filled with regular zebrafish systems water and 200 mg/l Tricaine. Using phase contrast, the GFP expressing embryos were then turned so the dorsal side of the spinal cord faced the microscope objective directly; this was not always possible, because even though the round sample is immobilized in the agarose, the embryo is freely floating inside the chorion. Zebrafish already develops significant pigmentation on the dorsal side by around 24 hpf. Hence, if the purpose was to watch stages past around 24 hpf, 0.1-0.2 mM PTU (Li et al., 2012) final was added to the growth medium about 12 hpf to prevent pigment formation and thereby keeping the zebrafish transparent; 0.1-0.2 mM PTU final was also added to mounting solutions and the observation chamber fluid during light-sheet microscopy.  Embryos older than 24 hpf were dechorionated, if not already hatched, and mounted in either 1-1.5\% agarose or in an uncoated FEP (Fluorinated Ethylene Propylene) tube with regular zebrafish system water. FEP having about the same refractive index as water (Weber et al., 2014). The FEP tube was briefly precleaned with 70\% ethanol and water prior to use. The same piece of FEP tube was used multiple times, minimizing material and cleansing needs.  The zebrafish can grow and extend in the FEP-tube and was used for imaging zebrafish up to 40 hours continuously; placing the fish head upwards improved survival. Adjustment of significant drift caused by the soft FEP tube was sometimes necessary; in comparison, little or no drift occurs for agarose mounted samples.

\eject

\noindent \textbf{2.3 Microscopy}

\noindent The SciLifeLab BioVis Zeiss Z.1 Light-Sheet microscope equipped with standard chamber (30 ml) for use with a 20 x observation objective (NA=1) and 10x illumination objective was used (Zeiss). Image acquisition was every 8${}^{th}$ minute, typically x,y 1920x1920 pixels, or about 500  \textgreek{m}mx500 \textgreek{m}m, zoom 0.8-1.4. GFP was visualized at 488 nm at approximately 2-8\% laser power, with pivot scanning, at 26-29$\mathrm{{}^\circ}$C. The early spinal cord is not more than about 100x100 \textgreek{m}m (widthxdepth) but the x,y,z image field was chosen big to accommodate for drift, movement and growth during the long term recordings. Deep z-stacks of the spinal cord, typically 200-700 slices with each slice having a thickness of 0.19 \textgreek{m}m, were made to cover both above and below the region of interest (ROI) during growth. To increase data amounts acquired per recording sessions, several larvae were frequently mounted at the same time in the microscope; also because the curved form of the larval spinal cord limits the field of observation and the larvae may change position significantly (due to growth inside the chorion) at this stage of development. For such multi-embryo mounting, the Multiview menu in the microscope operating software was used to fix the x,y,z imaging coordinates for each embryo for each loop of the time lapse. Only one view was taken for each embryo. Multiview was, hence, not used in attempt to improve the point spread function (PSF), and thereby resolution by imaging the same volume of the specimen from several angles; Multiview applied this was did not seem to improve resolution in this transparent zebrafish specimen; moreover, it gives more files to process, becoming impracticable for long term acquisitions. Adjustment of the light-sheet was best done manually, using a few days old zebrafish with fluorescent structures of the kind to observe; alternatively, the light-sheet adjustment wizard was used. Frequently, a narrower light-sheet (manually adjustable in Advanced settings) was used, than set by the software, as it gave sharper images of the submicron structures observed in this study.

\vspace{\baselineskip}

\noindent \textbf{2.4 Micro injections}

\noindent Needles with a tip opening of about 10-15 \textgreek{m}m were freshly made by manually breaking with forceps (Dumont{\circledR}, size 5) a drawn-out glass capillary with filament (1 mm diameter, Harvard Apparatus{\circledR} \#BS430-0032). Ejection pressure calibrated to give a roughly 1 nl drop at each ejection prompt. Embryos at the 1- to 2- cell stage were immobilized by letting them adhere to the long edge of a standard microscope slide positioned in a 15 cm petri dish and excess water sucked away (holding about 50 embryos). The needle mounted on a micromanipulator and approached to the chorion, poked through the chorion and, once inside the yolk, pressure ejection activated and the needle immediately withdrawn. Morpholino stocks were at 1 mM in Millipore{\circledR} water, about 8 ng/nl, and if diluted this was done just prior to the experiments. 5'-3' sequences for the two zRobo3, Robo3var1 AAAGTCCTGCGAAACTCCATCAGCC, Robo3var2 CGTCTTTATCAGGTAACGCAGCATC, based on the Genebank sequence data AF337036 and \\
AF304131, respectively (Challa et al., 2005); the Gene Tools standard control morpholino was used as control.

\vspace{\baselineskip}

\noindent \textbf{2.5 Single cell analysis}

\noindent From 2 dpf GFP expressing zebrafish, the yolk bag was removed manually with forceps (Dumont 5). The zebrafish head was cut off with forceps at the junction between the spinal cord and the hind brain. The remaining full length spinal cord cut in half, giving rise to two pieces of about equal length. The top (rostral) part was with a pipette transferred to a drop of 0.25\% trypsin-EDTA solution and with forceps teared briefly somewhat apart; then transferred to a 1.5 ml Eppendorf tube filled with 0.25\% trypsin-EDTA, which had been preheated in a heating block at 31$\mathrm{{}^\circ}$C with water-filled sample holders. 2 to 3 spinal cord pieces/tube were used. Digestion was for about 30 minutes and the tubes were manually shaken gently every second minute or so, both to monitor the digestion and to improve the exposure of the spinal cord to the digestion mix. To stop the reaction, one volume defined trypsin inhibitor with 4 \% final Fetal Calf Serum added was added at room temperature; each digestion mix tube hence doubled in volume to 3 ml (2\% final FCS); if several digestions were carried out, they were pooled at this stage. The pooled mix was then promptly poured over first a 40 and then a 20 \textgreek{m}m filter to remove large debris. The mix then transferred to 1.5 ml tubes and centrifuged at 380 g, 5 min at 10-20$\mathrm{{}^\circ}$C. The supernatant removed, and the pellets one by one resuspended in the same 100 \textgreek{m}l PBS at room temperature and analyzed by FlowSight{\circledR} (amnis{\circledR}) at medium to high flow rate.

\vspace{\baselineskip} 

\noindent \textbf{2.6 Data analysis and calculations}

\noindent Light-sheet files were typically 0.1-1 terabytes (tb) and were manually trimmed using the Subset function in the used Zeiss Zen (black edition) software (RRID: SCR\_013672); if possible, cutting away parts of the x coordinate and the z-coordinate, when containing no information. The z-coordinate was further trimmed so only slices containing the commissural axons (without the cell bodies) were left, as the strong emission from the cell bodies disturbed subsequent processing. For commissural midline studies, a maximum intensity projection (MIP) was then performed, a step that reduces the file size significantly; given that the floor plate is flat and the specimen is transparent there was virtually no visual difference between the MIP projected and the fully trimmed z-stack image file. The obtained MIP file was then used for manual measurement of axonal growth; however, if a specific structure is studied, even higher image quality may be obtained by selecting, at each time point, the best slice of the ROI (exemplified in Figure 4); if necessary, analysis of the full z-stack slice by slice was used similarly to ascertain that the same axon was followed over time. For three dimensional (3D) display and rendering, files were trimmed further in Imaris{\circledR} (RRID: SCR\_007370; Bitplane{\circledR}), which was also used to render the live recordings as rotatable 3D live recordings when necessary to ascertain that the same axon was followed over time (e.g. Movie 2 and 3). Igor Pro (RRID: SCR\_000325; WaveMetrics{\circledR}) and Excel (Microsoft{\circledR}) were used to analyze and plot the neuronal growth data. Movies were made with ImageJ or Fiji (RRID: SCR\_003070, RRID: SCR\_002285) open-source software (directly on the Zeiss light-sheet czi-format image files, or after first having exported each time point as a full resolution 16 bit Tif file) or Imaris (czi files of the z-stack converted to ims files); motion correction with Stack Reg (ImageJ or Fiji) or drift correction (Imaris). Deconvolution was done on the raw z-stack using the Huygens Software (RRID: SCR\_014237) of Scientific Volume Imaging{\circledR} with their light-sheet (SPIM) software module and auto-wizard generated settings. Stitching of the spinal cord was made using Arivis{\circledR} or the open-source XuvStitch (RRID: SCR\_005894; www.xuvtools.org), giving comparable results; each portion of the spinal cord consisted of about 600 images in a z-stack (both Arivis and XuvStich require the stacks to be stitched to have the same number of z-slices). The stitched image was subsequently reconstructed and analyzed in Imaris using clipping plane and surface reconstruction tools. The FlowSight data were analyzed with the IDEAS{\circledR} software (amnis). In the morpholino microinjection experiments, the number of midline crossing commissurals was counted manually and the number of midline crossing commissurals/\textgreek{m}m length spinal cord used in the statistical test. The angle of growth cone wandering was measured frame by frame using the Graphics options in the Zeiss Zen (black edition) freeware. Statistical tests were two-tailed t-test for two independent population means. Error bars are standard deviations (sdev) except in the midline crossing plot where standard errors of the mean (sem) are displayed.

For the midline crossing plot (Figure 1b), each axon's length was measured at each time point (as previously described for microtubules (Andersen et al., 1994)) and the instantaneous growth rate (y) calculated by comparison with the axon's length at the immediate previous time point, i.e. \textgreek{D}length/\textgreek{D}time=(t2 length - t1 length)/(t2-t1) measured in \textgreek{m}m/min, and the physical position (x) of the growth cone on the floor plate at that time point noted. The purpose of using the instantaneous growth rate is to look for variations in the commissural growth rate at as high as possible temporal resolution and then correlate that information with the physical position of the growth cone on the floor plate at that time. The time resolution on the instantaneous growth rate is 60/8 or 7.5 measurement points/hour, given that an image frame was taken every 8th minutes (i.e. \textgreek{D}time=t2-t1=8 min). For each calculated instantaneous growth rate (\textgreek{D}length/8min), the physical position of the growth cone was expressed by the fraction of the floor plate crossed at the given time. A number between 0 and 1, and the exact midline has then the value of 1/2. The fraction arises by measuring how far the commissural growth cone has grown at the given time point and dividing it by the total distance to cross the floor plate. Hence, a normalization of the total distance grown compared to the total distance to grow, and calculated independently for each neuron. Thus, the N=11 axons (those shown in Figure 1a, Movie 1) were analyzed at each 8${}^{th}$ minute time point, giving rise to a total of n${}_{tot}$=101 measurements of instantaneous growth rate (\textgreek{D}length/8min), and for each of these 101 time points there was then a corresponding fraction of the floor plate crossed, giving n${}_{tot}$=101 (x,y)= (fractional crossing, \textgreek{D}length/8min) coordinate sets. Regardless of from which axons the growth rates were derived, the 101 (x,y) coordinate sets were sorted according to their growth rate (y) into growth rate bins with a width of 0.1 \textgreek{m}m/min (thus with the bins representing growth rate intervals of  0-0.1, 0.1-0.2, 0.2.-0.3, 0.3-0.4, 0.4-0.5, 0.5-0.6, 0.6-0.7, 0.7-0.8, 0.8-0.9 \textgreek{m}m/min). The 101 \textgreek{D}length/8min growth rate measurements were distributed into the following bins: 0.05(n=22), 0.15(n=14), 0.25(n=22), 0.35(n=16), 0.45(n=10), 0.55(n=9), 0.65(n=6), 0.75(n=1), 0.85(n=1) \textgreek{m}m/min  (the two bins `0.75' and `0.85' were not used subsequently as these instantaneous growth rates were only observed n=1 times). For example, if at a particular time a commissural could be observed growing at an instantaneous growth rate (y) of \textgreek{D}length/8min=0.33 \textgreek{m}m/min at a normalized fractional midline crossing position (x) of 0.47, this coordinate set would be placed into the y=`0.35 \textgreek{m}m/min'-bin, used for all coordinate sets having growth rates in the interval between 0.3-0.4 \textgreek{m}m/min, and since the bin's midpoint-value ((0.3+0.4)/2=0.35) is used for averaging of the bins growth rate-interval value for the subsequent graphical display (Figure 1b); here, n=16 of the total 101 growth rate points were observed to be in this 0.3-0.4 interval.  Another example could be a growth rate of y=0.62 m/\textgreek{m}min at a crossing of x=0.58 which would then be placed in the y='0.65 \textgreek{m}m/min'-bin that represents all growth rates observed to be in the interval from 0.6-0.7 \textgreek{m}m/min; here, n=6 of the total 101 growth rate points were observed to be in this 0.6-0.7 interval. Regardless of which of the N=11 axons the growth rates were derived from, all the 101 (x,y) measurements were this way distributed into their corresponding growth rate bin. Hence, all points in each growth rate bin had an associated normalized fractional midline crossing point (given mentioned 101 (x,y)=(fractional crossing, \textgreek{D}length/8min) coordinate sets).
 
Then, in principle, it were possible that growth rates in a given growth rate bin could stem from any fractional midline crossing point, i.e. from any position (x) during the crossing of the floor plate. To test this, for each growth rate bin the average normalized fractional crossing value with sem was then calculated and plotted against the bin \textgreek{D}length/min growth rate value (y) (see Figure 1b); it is seen from the plot that slow growth rates cluster close to the midline, which has the normalized fractional floor plate crossing value (x) of 1/2 (illustrated in the Graphical Abstract in Supporting Information, stippled blue line). Hence, growth rates on the floor plate do not vary randomly but tend to be lower near the midline. A quadratic curve fit of the plotted data rendered a fitting polynomial (y = 3.57 -- 13.59x + 14.05x${}^{2}$), used to draw the stippled curve in Figure 1b. Extrema of a defined function can be found by differentiation. Thus, by differentiating the fitting polynomial, it is seen that there is an average minimum in the growth rate \textgreek{D}length/min (y) at the normalized fractional floor plate crossing point of x = 0.484, where the midline is at the exact midpoint with a value of 1/2. 

In the plot inserted in Figure 1b, the location of the midline is now at 0.0. This has been obtained by subtracting 0.5 from the x-values values. Plotted is then the growth rate value (y) versus the \textit{absolute} x-value, such that this graphical representation does not distinguish whether the growth cone still has a distance left to reach the midline (negative x) or has already crossed the midline and is at a distance away from the midline (positive x).\eject 

\noindent \textbf{3. RESULTS}

\vspace{\baselineskip}

\noindent \textbf{3.1 Dynamics of commissural axons crossing the ventral floor plate in real time}

\noindent To obtain live real time \textit{in vivo} recordings of axons crossing the midline in the zebrafish spinal cord by light-sheet microscopy, a transgenic Tg(dmrt3:GFP) zebrafish that express GFP under control of the dmrt3 promotor was used in this study (Satou et al., 2013). The first observable GFP signal was 10 hours post fertilization (hpf) in the nostral region of the forebrain (Satou et al., 2013), and 13-16 hpf was the earliest time points where a strong GFP signal in the spinal cord could be observed with a regular stereomicroscope and which is just prior to the outgrowth of pioneer axons in the spinal cord (Geldmacher-Voss et al., 2003; Kimmel et al., 1994; Papan and Campos-Ortega, 1994; data not shown). The strong GFP expression made it likely that it would be possible to follow early commissural axonal wiring in the spinal cord that begins around 16 hpf (Bernhardt et al., 1990; Drapeau et al., 2002; Kimmel et al., 1994; Kuwada et al., 1990a). 

When embryos aged 16-28 hpf were imaged, commissural axons could be observed to cross the floor plate at a growth rate of about 18.1 \underbar{+} 6.4 \textgreek{m}m/h (N=12 axons in 4 different zebrafish, representing 20.8 h of total axonal growth). It was therefore endeavored to observe the first commissural axons that start crossing the floor plate around 16 hpf. Looking at the ventral midline of the rostral half of the spinal cord, in one embryo and starting about 18 hpf, commissural axons could be observed to utilize around 2 to 4 hours to cross the 55 \underbar{+} 6 \textgreek{m}m wide ventral floor-plate and reach the contralateral side; the average growth rate during this journey was 0.35 \underbar{+} 0.14 \textgreek{m}m/min or about 21 \underbar{+} 8.4 \textgreek{m}m/h (N=11 axons in 1 zebrafish, representing 18.5 h of total axonal growth; Movie 1, Figure 1; Supporting Information Movie S1).

\vspace{\baselineskip}
\begin{center}
\includegraphics*[width=4.56in, height=4.78in, trim=0.90in 3.32in 0.77in 0.89in]{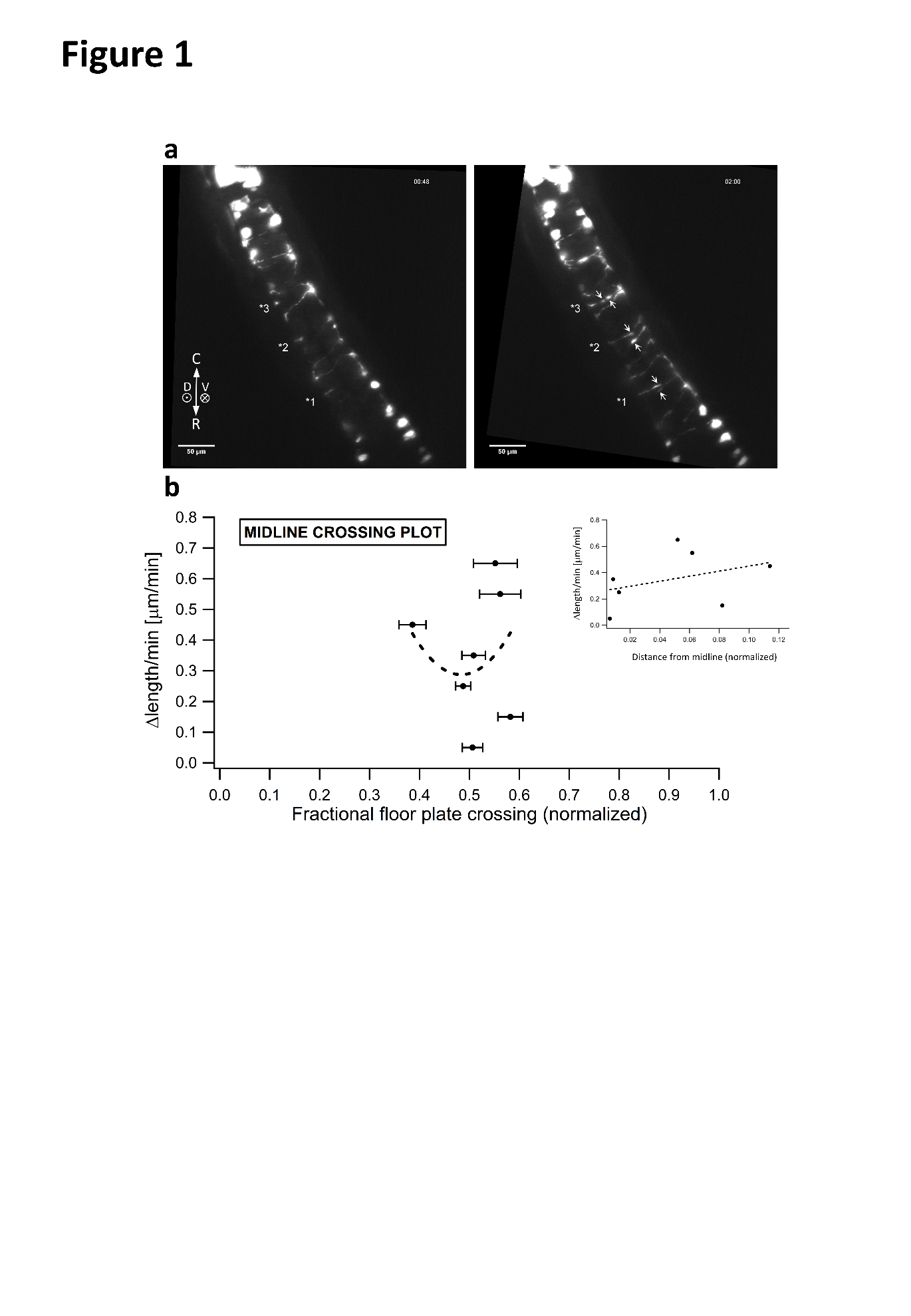}
\end{center}

\noindent \begin{small} \textbf{Figure\_1} Legend. Real-time observations showing commissural axons crossing the midline in temporal and spatial synchrony (see accompanying Movie 1, 2 and 3). \textbf{a}) Left, commissural axons at three different points (each (\textit{1}-\textit{3}) marked by an asterisk (\textit{*}), \textit{t} = 48 min) start crossing simultaneously at 18 hpf from either side of the floor plate (Movie 1); right, the commissurals have reached the midline (growth cones of the three commissural pairs are marked by six arrows, \textit{t} = 2 hr); the compass shows spatial directions and Rostral (R) is towards the bottom, Caudal (C) top, Dorsal (D) is above ($\mathrm{\odot}$) and Ventral (V) below (\circletcross) the field of view, bar 50 \textgreek{m}m. (Movie 1). \textbf{b}) Midline crossing plot (derived in section 2.6), quantifying that the midline crossing commissural axons (shown in (a), Movie 1) tend to grow slower in the area close to the midline (N=11 axons in 1 zebrafish, errors are sem); this is illustrated by the stippled quadratic trendline, indicated as a guide for the eye, suggesting a minimum growth rate very close to the exact floor plate midline (located in the plot at the fractional floor plate crossing point of 0.5). \textit{Insert}: the same data, here plotted as the growth cone speed against the absolute fractional distance to the midline - thus, the midline is here located at x = 0; the stippled linear trendline is indicated as a guide for the eye, and this representation also illustrates that the commissural axons tend to grow slower in the vicinity of the midline. (Movie 1).\end{small}  

\vspace{\baselineskip}

Moreover, along the length of the spinal cord, a striking synchrony of commissural axonal pairs crossing the midline with bilateral symmetry in adjacent segments was observed. Hence, on either side of the midline, three pairs of commissurals are seen crossing in temporal synchrony at a rostral-caudal interval of approximately 100 \textgreek{m}m (labelled with asterisks in Movie 1, Figure 1a, starting at t = 48 min). At this early developmental stage, 100 \textgreek{m}m is approximately the length of a segment in the rostral spinal cord. To address whether the growth rate of these early commissurals vary during the passage of the floor plate, the growth rate data were analyzed graphically in the Midline crossing plot that reveals an experimental average minimum in the growth rate at a fractional-crossing close to the exact midline crossing point of 1/2 (Figure 1b, derived in section 2.6; Movie 1). In the vicinity of the midline choice point, growth cones appear to become bigger in cross sectional area as compared to before the midline. The decreased growth rate and increase in size are seen to be reversible (Figure 1b, Movie 1).

\vspace{\baselineskip}
\begin{center}

\noindent                                                                \includegraphics*[width=1.90in, height=1.90in]{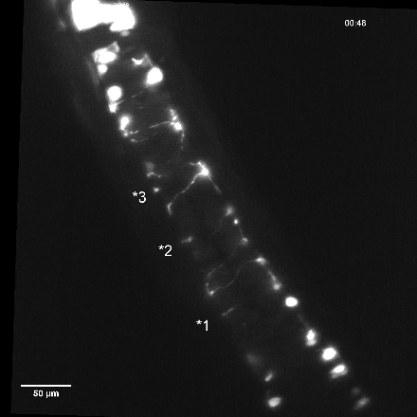}

\end{center}

\noindent \begin{small} \textbf{Movie\_1} Legend. (See Supporting Information). Commissural axons crossing the floor plate (about 18 hpf). Viewing the movies by scrolling back and forth frame by frame is most informative (possible e.g. with Windows' `Movies \& TV' application). The spinal cord in a maximum intensity projection, viewed from the dorsal side (from straight above). Three marks indicate pairs of commissural axons that are crossing simultaneously (*1, *2, *3; starting at t=48 min) from either side and at the same time (notice the slight rostral-caudal temporal shift in axonal advancement, reflecting the rostral-to-caudal gradient in developmental progression). Rostral is towards the bottom in the Movie (Figure 1). In *2, a clearly visible enlargement and slowing down of the growth cones near the midline are seen. Caudally, notice an axon is taking a sharp turn caudally after crossing (clearly visible at t=6h08min-7h04min). The small arrows appearing in different locations in Movie 1 (e.g. at t=3h20min-3h52min, middle) are explained in the legend to Figure 4d'. This particular movie was used for the quantification in the Midline crossing plot (Figure 1b). The apparent rotation of the image field is due to sample drift during acquisition that is corrected prior to the display and causing the rotation. Length 7 hours 4 minutes in real time (see also Movie 2 and 3), bar is 50 \textgreek{m}m.\end{small}

\vspace{\baselineskip}

After crossing the floor plate, some commissural axons could be followed further (Movie 2 and 3; Figure 2). At the contralateral site, these axons could be observed ascending with spatio-temporal synchrony at an oblique asymptotic angle in three adjacent segments of the spinal cord, most likely toward the dorsal longitudinal fasciculus (DLF) (Movie 2; Figure 2a1-a3, three arrows). These same three commissurals may be distinguished during growth on the floor plate itself (Movie 3; Figure 2b, three arrowheads). The noticeable rostral-caudal temporal shift in the synchrony is consistent with the rostral-to-caudal gradient of developmental progression (Movie 2, Figure 2; a slight rostral-caudal temporal shift is also noticeable for the commissurals in Movie 1).

\vspace{\baselineskip}

\begin{center}
\noindent                           \includegraphics*[width=4.74in, height=1.50in]{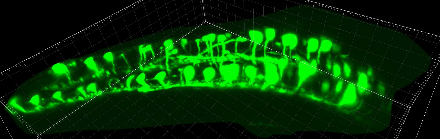}
\end{center}

\noindent \begin{small} \textbf{Movie\_2} Legend. (See Supporting Information). Axons growing in spatio-temporal synchrony at an obliquely ascending angle after crossing the floor plate (about 28 hpf). Same sequence as Movie 3. After having crossed the floor plate, three commissural axons in three different segments are growing in spatio-temporal synchrony at an obliquely ascending asymptotic angle most likely to join the DLF (Figure 2a; rostral is right). The cell bodies are clearly visible at the dorsal side, rostrally. Notice the slight rostral-caudal temporal shift in axonal advancement, reflecting the rostral-caudal gradient in developmental progression. Length 4 hours 48 minutes in real time, bar is 30 \textgreek{m}m.\end{small}

\vspace{\baselineskip}

\begin{center}
\noindent                                                                \includegraphics*[width=1.96in, height=1.80in]{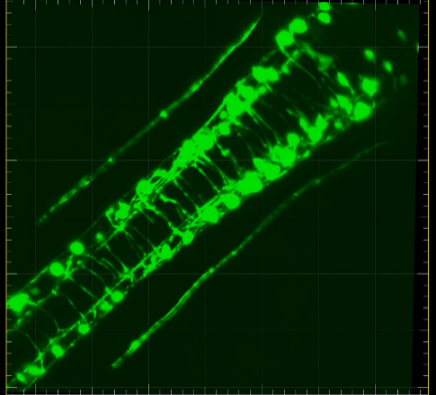}
\end{center}

\noindent \begin{small} \textbf{Movie\_3} Legend. (See Supporting Information). Commissural axons crossing the floor plate (about 28 hpf). This recording is the same as shown in Movie 2, where the view is tilted onto its side, here seen looking down from straight above the dorsal side (rostral is right). In this perspective, apart from many commissurals having already crossed or still crossing the floor plate, the three axons ascending in Movie 2 at an oblique asymptotic angle once the floor plate crossed (Figure 2a, arrows 1-3),  can here be observed crossing the floor plate from right to left (Figure 2b, arrowheads 1-3; correspondence was ascertained by rotatable 3D live rendering). Length 4 hours 48 minutes in real time, bar is 30 \textgreek{m}m.\end{small}

\vspace{\baselineskip}

\begin{center}
\noindent                           \includegraphics*[width=4.35in, height=5.64in, trim=1.06in 1.81in 1.01in 1.13in]{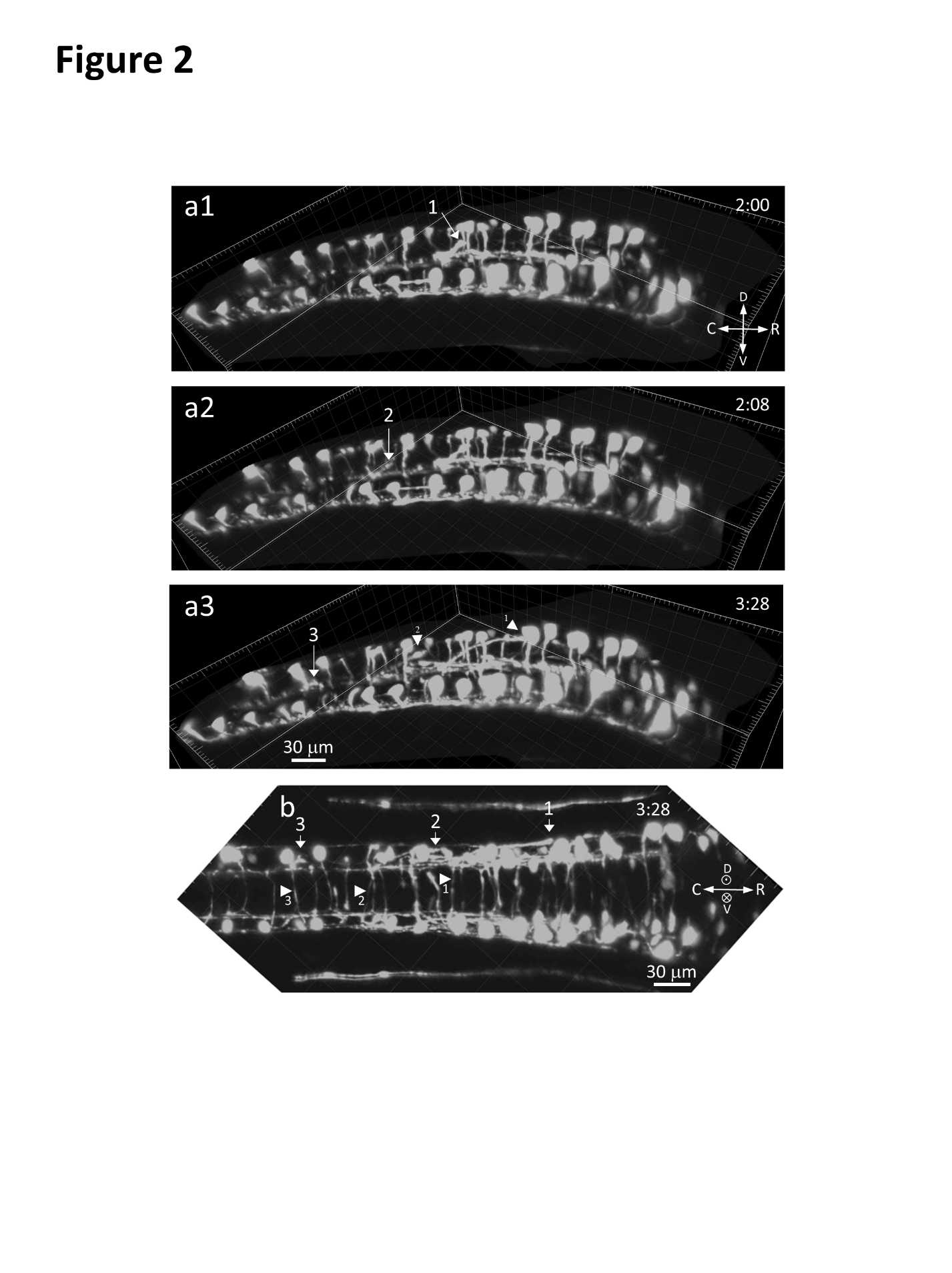}
\end{center}

\noindent \begin{small} \textbf{Figure\_2} Legend. After crossing the floor plate, obliquely ascending axonal growth contralaterally (28 hpf; see accompanying Movie 2 and 3). \textbf{a1}-\textbf{a3}) Having crossed the floor plate, three commissurals (indicated by arrows 1-3) are seen ascending at an oblique asymptotic angle in three different segments and with both spatial and temporal synchrony, most likely to join the dorsal longitudinal fasciculus (DLF), compass shows rostral (R) is to the right, time in hours:min in right corner, bar 30 \textgreek{m}m, (Movie 2). \textbf{b}) Same recording as in a1-a3 (dorsal (D) is above ($\mathrm{\odot}$) the field of view), where these 3 axons (arrows 1-3) are seen having crossed the floor plate (their shafts are indicated by the tip of arrowheads 1-3, all pointing to the right; correspondence was ascertained by rotatable 3D live rendering), compass shows directions, bar 30 \textgreek{m}m, (Movie 3). (Movie 2 and 3).\end{small}

\vspace{\baselineskip}

In the presented movies, the dorsal GFP-expressing cell bodies have been removed from the display because their intense GFP staining interferes with observations of the growing axons on the ventral floor plate. To illustrate this, Movie S2 (Supporting Information, Movie S2) shows an example with cell bodies. The axon starts growing out from the dorsally located cell body down in the ventral direction toward the ventral floor plate. Then crosses the ventral floor plate and on the contralateral side turns to join most likely the DLF (Movie S2).

Interestingly, almost no guidance errors were observed in the recordings in this study from more than 10 different zebrafish, representing more than 100 axons. An error is here defined as a growth cone straying at least 10 \textgreek{m}m off, measured orthogonally, the normal pathway (Hutson and Chien, 2002). Hence, in the recordings in this study it was never observed that a commissural axon would commence an ipsilateral path, for then to correct it as an error, and continue with a midline crossing growth path across the ventral floor plate. In converse, but only on one occasion and that in the future hindbrain, an ipsilateral axon was observed growing towards the floor plate midline as if a commissural (Movie S3a and S3b), for then to correct its path by rapidly shrinking back upon reaching the midline and continue the correct caudally directed ipsilateral path, most likely towards the medial longitudinal fasciculus (MLF; Bernhardt et al., 1990; Drapeau et al., 2002; Kimmel et al., 1994; Kuwada et al., 1990a). These observations suggest that axonal guidance errors of this magnitude (Hutson and Chien, 2002) are extremely rare events during the observed stereotyped spinal cord wiring \textit{in vivo}, that is, spinal cord wiring seems error-free. 

\vspace{\baselineskip}

\noindent \textbf{3.2 Growth cone wandering and bilateral filopodia and neurite shaft interactions }

\noindent The light-sheet technique enables high resolution acquisition along many hundred microns long regions that may then be viewed at variable zoom rates to display different information, such as the spatio-temporal synchrony at moderate zoom shown in Movie 1, 2 and 3. At higher zoom (Movie 4, 5 and 6; unlabeled Movie S4, S5 and S6, Movie S7; Figure 3 and 4), transiently slowing down and increasing the growth cone size was observed to be a generic guidance response of the commissurals. Not just as observed at the midline guidance point of the ventral floor plate (Movie 1, Figure 1b) but also at the contralateral guidance point (Movie 4 labels N1 (midline) and N2 (contralateral); Figure 3).

\vspace{\baselineskip}

\begin{center}
\noindent                   \includegraphics*[width=4.86in, height=6.50in, trim=0.71in 2.31in 0.86in 0.58in]{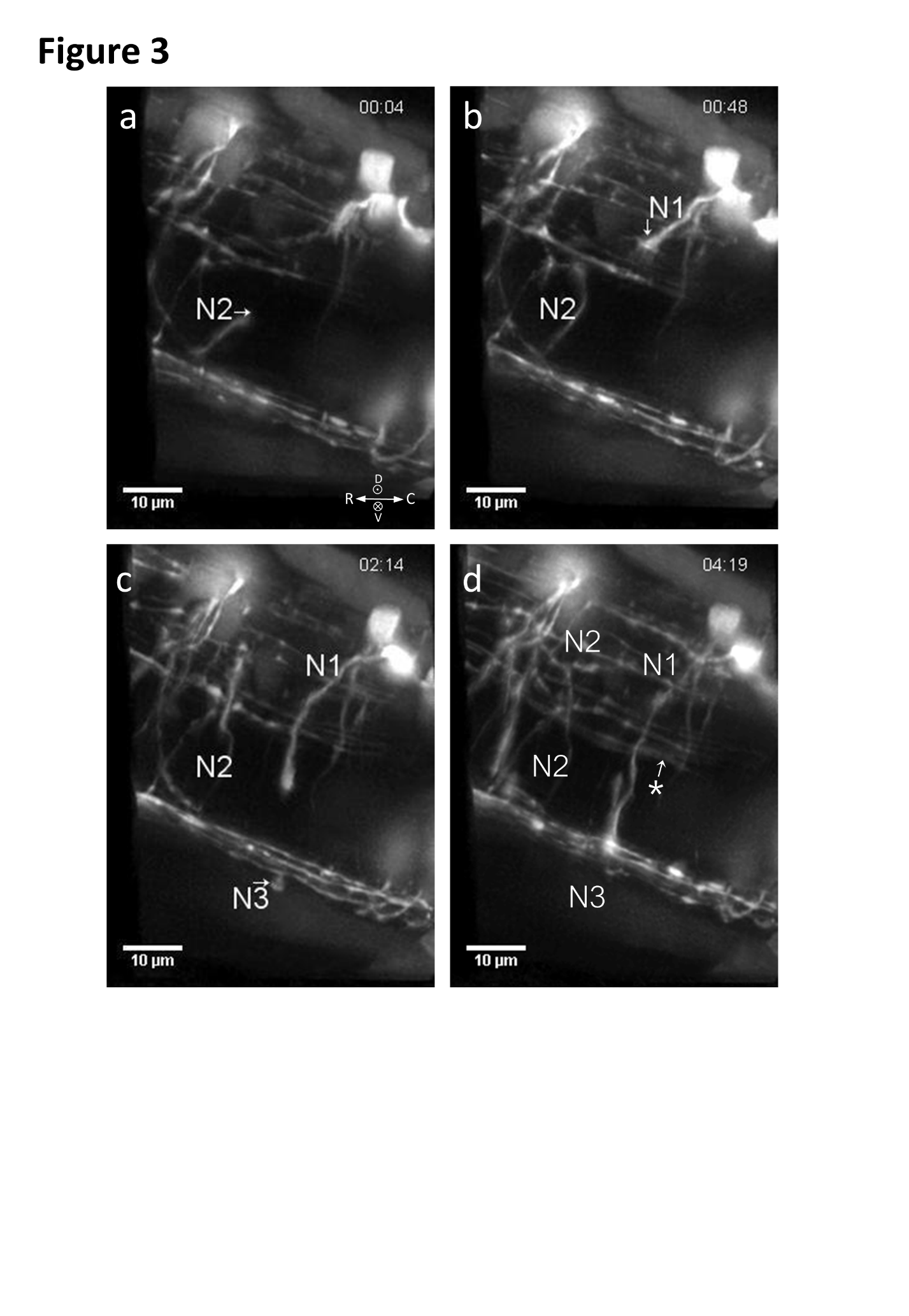}
\end{center}

\noindent \begin{small} \textbf{Figure\_3} Legend. Filopodia dynamics and growth paths of three midline commisurals and one ipsilateral axon (see accompanying Movie 4). Here, the already high-resolution raw light-sheet image recordings have been deconvolved, improving the resolution from about 800 to 600 nm, estimated by a full-width half maximum (FWHM) measurement of the Gaussian signal, as well as the signal-to-noise ratio. \textbf{a}-\textbf{d}) \textit{N1}, cross from the contralateral side toward the viewer, then turns right (caudally); \textit{N2}, cross the floor plate, then bifurcates on the contralateral side (at N2 in \textit{d} it has already bifurcated); \textit{N3}, this axon wraps around the ipsilateral axonal bundle (on the ipsilateral side) on the outside, for then to make a sharp turn onto the floor plate and cross to the contralateral side; \textit{*}, marks the growth cone of an ipsilateral axon that in Movie 4, at t=4h19min, can be seen extending fast rostral to caudal at the level of the contralateral ventral floor plate. Time in hours:min in right corner, compass shows directions, bar 10 \textgreek{m}m. (Movie 4).\end{small} 

\vspace{\baselineskip}

\begin{center}
\noindent                                                               \includegraphics*[width=1.46in, height=1.97in]{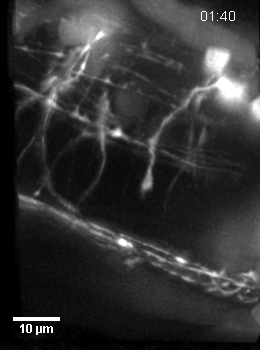}
\end{center}

\noindent \begin{small} \textbf{Movie\_4} Legend. (See Supporting Information). High resolution deconvolved sequence shows filopodia dynamics and growth paths of three commissural and one ipsilateral axons (Figure 3). Viewing the movies by scrolling back and forth frame by frame is most informative (possible e.g. with Windows' `Movies \& TV' application). \textit{N1}, grows down from the dorsally located cell body towards the floor plate, cross then from the contralateral side towards the viewer, slows down noticeably at the midline, then turns right (caudally) while extending multiple filopodia; \textit{N2}, cross the floor plate and then bifurcates on the contralateral side (commencing at t=2h28min); \textit{N3}, this axon wraps around the ipsilateral axonal bundle (on the ipsilateral side) for then to make a sharp turn onto the floor plate and cross to the contralateral side, where it stalls as its growth cone enlarges and searches, and which likely is causing the shaft bends seen close to the growth cone at t=7h21min-8h00min; \textit{*}, on the contralateral side, an ipsilateral axon extending fast rostral to caudal at the level of the ventral floor plate (in this Movie, other ipsilateral axons may also be seen extending). Movie S4 is an unlabeled version of this movie. Length 8 hours in real time, bar is 10 \textgreek{m}m. \end{small} 

\vspace{\baselineskip} 

When axons are growing, the path of their growth cones often deviate slightly from that of a straight line which was also observed clearly in this study at the higher zoom. To quantify this wandering, or searching, the absolute angle made, at each time point, between a commissural growth cone and the straight line connecting the growth cone's starting to its ending position (the straight path) was measured. This gave a wandering variation of (7.23$\mathrm{{}^\circ}$ \underbar{+} 6.17$\mathrm{{}^\circ}$)/8 min that appeared random (n=268 time points of 8 minutes, measured on N=18 axons in 4 different zebrafish, representing 34 h of commissural growth). The investigator induced error on this angle measurement was estimated empirically to around 1.22\underbar{+}0.75$\mathrm{{}^\circ}$ (n=14). Because all labelling is monochromatic green (GFP) in this study, the ipsilateral population of axons was not evident to follow at low magnification. At the higher zoom, they could be observed to grow at around 69.6 \underbar{+} 20.7 \textgreek{m}m/h (N=13 axons in 6 different zebrafish, representing 28.3 h of total axonal growth) and some were even observed to grow as fast as 120 \textgreek{m}m/h (Movie S8). Hence, ipsilateral axons appear to extend about three to six times faster than the commissural axons (Movie 4, 5, S8; Figure 3 and 4). The ipsilateral axonal growth cones, in comparison with the commissurals, grow straight with little wandering of (3.71$\mathrm{{}^\circ}$ \underbar{+} 3.83$\mathrm{{}^\circ}$)/8 min (n=136 points, measured on N=8 axons in 4 zebrafish, representing 21.4 h of total axonal growth), also considering their three to six times higher growth rates. The few abrupt changes observed in the ipsilateral growth cone direction seemed not random but caused by encountered obstacles (e.g. shafts of axons growing from the dorsal cell body towards the ventral floor plate). The wandering behavior is an intrinsic property of advancing growth cones (due to the ongoing molecular dynamic events inside the growth cone (Andersen and Bi, 2000)) and a read-out of the growth cone angle is obtainable at each time point. Thus, even if the image acquisition frequency was increased, giving higher time resolution, these growth cone wandering angle estimates would likely not change.

\vspace{\baselineskip}

 Interestingly, the commissural shafts could often be observed making localized \\ sinusoidal-shaped bending movements on the floor plate, both during and after crossing (Movie 5 and 6, Figure 4). They appear from the recordings to be mostly caused by the stochastic transient actions of filopodia emanating from neighboring growth cones that are visible at higher zoom and having lengths of up to more than 10 \textgreek{m}m and which are very responsive to the environment; the filopodia seems to collide transiently with shafts in the vicinity, and appear thereby to cause a local and reversible displacement of the shaft in the direction of the growth cone of the colliding filopodia, and appear thereby to cause the observed transient shaft bends (Movie 5 and 6, Figure 4), as also proposed by Moon and Gomez (2005). These transient bends are not, unlike growth cone wandering, caused by intrinsic properties of the axons. Therefore, quantifying the occurrence of shaft bends is strongly dependent on the image acquisition frequency. A higher image acquisition rate will increase the probability of observing the bends. At the used acquisition rate, an estimate of (0.13\underbar{+}0.14)/8 min axonal shaft bends was obtained (n = 235 time points, N=16 axons in 3 different zebrafish). Representing about 1\underbar{+}1 commissural axonal shaft bends per hour per commissural axon. Looking at the complementary scenario of a filopodia from an observed growth cone making contact with and apparently causing the bend of a neighboring commissural shaft, an occurrence of (0.044\underbar{+}0.081)/8 min contact-bending events were measured (Movie 5 and 6, labels 2, 3 and f; Figure 4d,d'); or approximately 1 contact-bending event every 3 hours \underbar{+} 1.7 hours. This is a three times lower frequency than aforementioned axonal shaft-bending frequency. That is to be expected because the GFP-labelled neurons are a minor fraction of all growing commissural and ipsilateral axons. Hence, only GFP-labelled filopodia are visible in this study whereas all growth cone filopodia may potentially cause bending of neighboring shafts.

While the ipsilateral growth cones also had long filopodia, they appeared more collected than their commissural counterparts. From the obtained recordings, the ipsilateral shafts did not seem to display noticeable sinusoidal-shaped bending movements (Movie 2, 3, 4, 5, 6 and S8).

\vspace{\baselineskip}

\begin{center}
\noindent              \includegraphics*[width=5.20in, height=7.70in, trim=0.57in 0.74in 0.65in 0.76in]{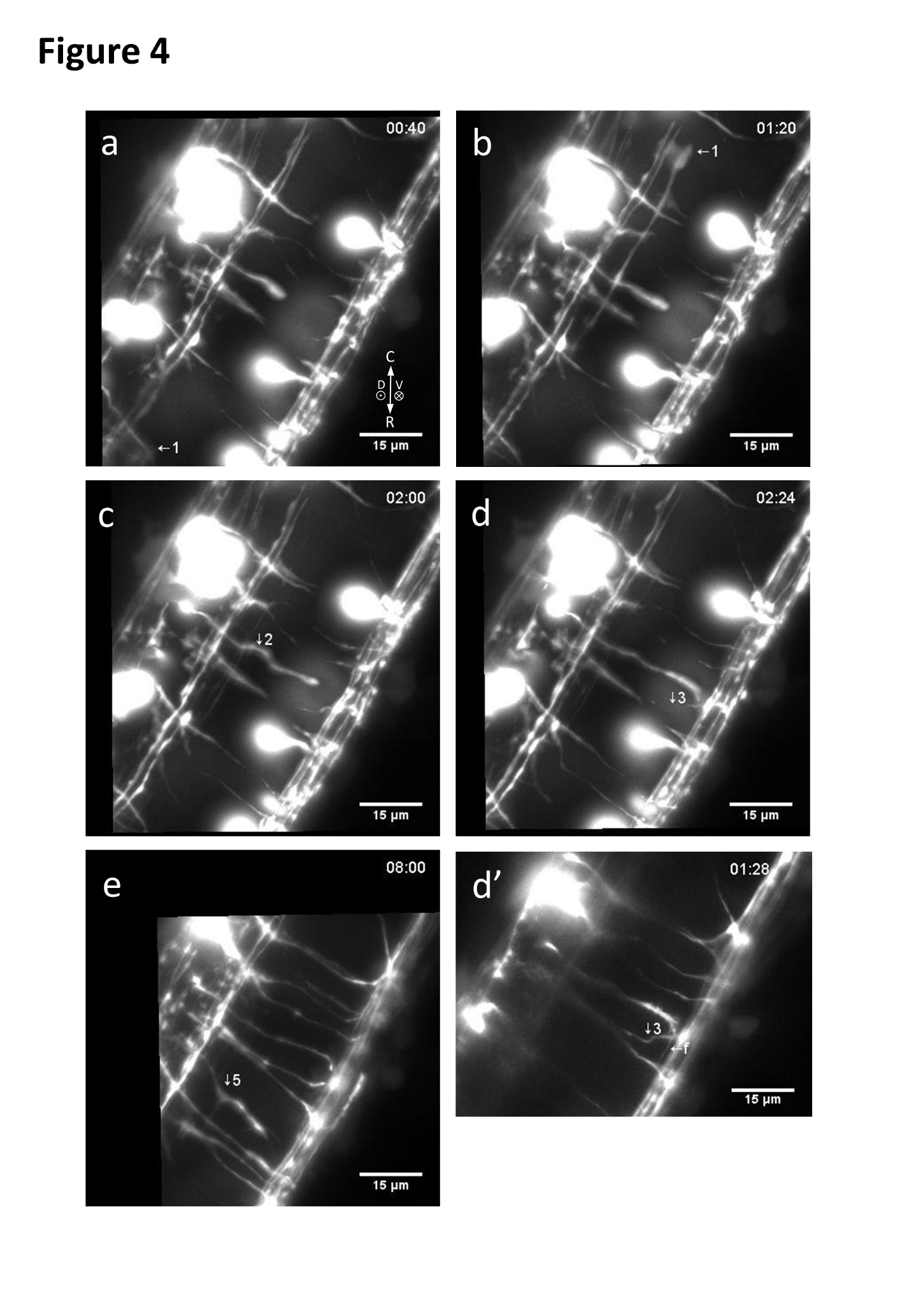}
\end{center}

\noindent \begin{small} \textbf{Figure\_4} Legend. Commissural axonal shafts and filopodia display bidirectional dynamics during floor plate crossing (see accompanying Movie 5 and 6). \textbf{a}-\textbf{b}) \textit{1}, an ipsilateral axon is here followed for 40 min showing no shaft bends; \textbf{c}-\textbf{e}) four examples of commissural sinusoidal-shaped shaft bends (labels \textit{2}-\textit{5}; \textit{4} occurs on the same shaft as \textit{2} and is only marked in the accompanying Movie 5, starting at t=5h44min; Movie 5); \textbf{d'}) This is the same area and recording as image d), showing how the filopodia (\textit{f}) from the growth cone of \textit{2} collides with a neighboring shaft \textit{3} and appears to cause it to bend. Right corner: time in hours:min, compass shows directions, bar 15 \textgreek{m}m. The difference in resolution between d) and d') is caused by different image handling. Hence, image (d) is a MIP of stacks (just as Movie 5) whereas (d') (Movie 6) only shows the one single slice-plane (0.2 \textgreek{m}m thick) of the stack where the object to observe (label 3) was clearest. As can be seen, the latter display option, while having better resolution by removing unwanted signal, has the drawback that the usable field of view is significantly decreased. The small arrows inserted in Movie 1 (e.g. at t=3h20min-3h52min, middle) are pointing at other examples of occurrences of sinusoidal-shaped shaft bends; they look smaller in magnitude than in Movie 5 and 6 because the zoom magnification is smaller in Movie 1. In actual size, the small arrows inserted in Movie 1 are twice as big as the arrows used in Movie 5 and 6. (Movie 5 and 6).\end{small}

\vspace{\baselineskip}

\begin{center}
\noindent                                                           \includegraphics*[width=1.96in, height=1.97in]{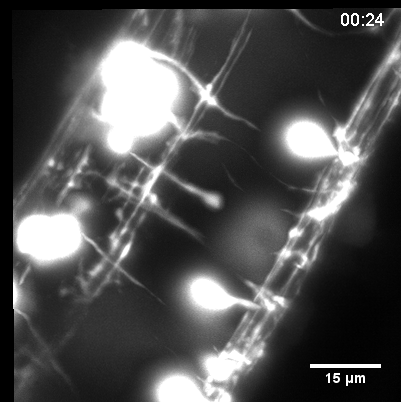}
\end{center}

\noindent \begin{small} \textbf{Movie\_5} Legend. (See Supporting Information). Commissural axonal shafts bend and wobble on the floor plate (Figure 4). \textit{1}, an ipsilateral axon extending straight and fast (other ipsilaterals can be observed in this movie); \textit{2}-\textit{5}, examples of commissural axonal sinusoidal-shaped shaft bends on the floor plate; Movie 6 is more focused on commissural \textit{3} only, and where a sinusoidal-shaft bend can clearly be observed and apparently caused by the collision of a filopodia emanating from the growth cone of the neighboring axon \textit{2}. Movie S5 is an unlabeled version of this movie. Length 8 hours 16 minutes in real time, bar is 15 \textgreek{m}m. \end{small} 

\vspace{\baselineskip}

\begin{center}
\noindent                                                  \includegraphics*[width=2.68in, height=1.99in]{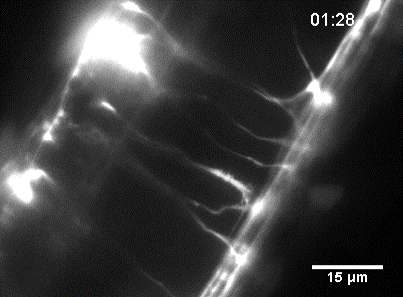}
\end{center}

\noindent \begin{small} \textbf{Movie\_6} Legend. (See Supporting Information). Commissural neurite shafts respond with sinusoidal-shaped bends upon contact by filopodia tips (Figure 4). Focused view of axon \textit{2} and \textit{3} in Movie 5. From t=1h20min to 1h44min, a filopodia of commissural \textit{2} appears to collide with the shaft of commissural \textit{3}, the shaft of \textit{3} appears then to be pulled in the direction of axon \textit{2}'s growth cone and this lateral displacement in turn appears to cause \textit{3} to make a reversible sinusoidal-shaped bend (Figure 4 d, d' at t=1h28min) that is resolved shortly thereafter (no longer observed at t=1h44min). Movie S6 is an unlabeled version of this movie. Length 2 hours 48 minutes in real time, bar is 15 \textgreek{m}m. \end{small} 

\vspace{\baselineskip}

\noindent \textbf{3.3 Probing factors regulating commissural midline crossing and single cell analysis}

\noindent Obtaining \textit{in vivo} recordings by light-sheet microscopy of commissural axons crossing the spinal cord midline at the floor plate was the primary rationale for this study. However, the accompanying important questions of molecules regulating guidance were considered, also to probe technically, whether it would be feasible to use molecularly perturbed zebrafish for the \textit{in vivo} light-sheet experiments. Robo3 receptors were knocked down by conventional translation blocking morpholino-antisense techniques. Morpholinos were chosen because they have already been applied to knock down Robo3 in zebrafish. Phenotypes for both Robo3 mutant zebrafish and translation blocking morpholino treated zebrafish have been reported (Burgess et al., 2009; Challa et al., 2005). In the Robo3 morpholino treated spinal cords, the number of commissurals per micrometer spinal cord that were observed to cross the floor plate was significantly reduced compared with the control (Figure S1b, p$\mathrm{<}$0.01; Supporting Information), in accord with previous data (Burgess et al., 2009). As expected, the real time observations showed that the commissural axons appeared to be avoiding the midline region of the ventral floor plate whereas ipsilaterally projecting axons were not affected (Movie S9). These results show light-sheet microscopy can be applied for \textit{in vivo} studies of molecularly perturbed zebrafish. However, following microinjections, a general cumbersome reduction in the number of both healthy and strongly GFP expressing zebrafish embryos and larvae was observed. Indeed, it would be timely to shift from morpholinos, affecting the whole embryo, to less toxic and more precise cell-specific gene editing tools such as the CRISPR-cas systems, as discussed by e.g. (Rojo et al., 2018; Schulte-Merker and Stainier, 2014). 

To address whether the commissurals consist of molecularly distinct axons, specific neuronal markers need to be generated, which could be accomplished by isolating individual neurons from the spinal cord. To that end, the rostral most half of the spinal cord was first digested with 0.25\% trypsin until a suspension of individual single cells was obtained. By FlowSight, each individual cell in the cell suspension was then imaged while flowing past a camera (Figure S1c, Supporting Information). Such single-cell approach may then be aimed at a subsequent single-cell omics analysis of other GFP-labelled neuronal populations in the spinal cord to gain a better understanding of the types of spinal neurons observed \textit{in vivo} (Andersen, 2001; Zeisel et al., 2015).\eject 

\noindent \textbf{4. Discussion}

\noindent Finally seeing both \textit{in vivo} and in real time the simultaneous commissural midline crossing events over extended lengths of the spinal cord is illuminating. In particular noticeable is the stereotyped spatio-temporal commissural growth synchrony (Movie 1, 2 and 3). The movies show the type of live organismal scale data that can be obtained by applying light-sheet microscopy and the zebrafish model system to the study of cellular developmental neurobiology.

\vspace{\baselineskip}

\noindent \textbf{4.1 Crossing the midline and wiring the spinal cord}

A minimum in commissural growth rate was observed very close to the midline (Figure 1b; Graphical Abstract in Supporting Information (stippled blue line); Movie 1, 2 and 3). A decrease in growth rate is expected at a choice point (Andersen, 2005; Bonner et al., 2012; Goodhill, 2016; Michalski et al., 2013; Rosoff et al., 2004). At the midline, both a decrease in growth rate and an increase in growth cone size have been reported previously (Bak and Fraser, 2003; Mason and Wang, 1997; Moon and Gomez, 2005; Myers and Bastiani, 1993b). Confirming previous reports, Robo3 is shown here to be a positive guidance factor for commissural axons; ipsilateral axons, as expected, appear guided independently of Robo3 (Figure S1b, Movie S9). How ipsilateral axons are guided remains an open question that is considerably less studied than commissural guidance. A plausible role for diffusible ipsilateral cues derives from recent data (Kastenhuber et al., 2009; Kim et al., 2014; Paix\~{a}o et al., 2013; Sakai and Kaprielian, 2012; Schweitzer et al., 2013).

Earlier static wiring studies classified the commissurals based on morphology. Thus, the CoPA (Commissural Primary Ascending) are here the commissural axons observed crossing perpendicular to the midline (Movie 1 and 2), and continuing an ascending growth at an oblique angle toward the DLF on the contralateral side (Movie 2). The CoSA (Commissural Secondary Ascending) are seen crossing the floor plate later (Movie 2 and Move 3). In Movie 4, axon N2 bifurcates upon reaching the contralateral side, identifying it as a commissural bifurcating longitudinal (CoBL (Hale et al., 2001); Figure 3a-d). A minor population crossing the midline obliquely at 1-1.5 dpf, and slightly more dorsally, may likely be identified as Commissural Local (CoLo) interneurons (data not shown; Satou et al., 2009; Shah et al., 2015).

An elucidation of the full wiring paths and origins (e.g. some axons may originate from the developing brain (Movie S3a and S3b) and not the spinal cord), synaptic partners and functions of the axons studied here require more experiments. A combined \textit{in vitro} and \textit{in vivo} approach will be required to characterize both the molecular and cellular properties of the different neurons observed in the spinal cord (Andersen, 2001). A significant step forward in that characterization will be the establishment of specific molecular neuronal markers. This may be accomplished by applying single cell analysis (Zeisel et al., 2015) of the FP-labelled neurons as illustrated and initiated here by the single cell FlowSight{\circledR} experiments (Figure S1c). A number of FP-spinal cord neuronal transgenic zebrafish lines have been described already (Satou et al., 2013), and the single cell approach may be applied to all spinal cord neurons to gain further knowledge about their specific localization, properties and functions. In comparison, for example, a fixed number of three primary motoneurons per hemisegment have been described (CaP, MiP, RoP; Lewis and Eisen, 2003).

\eject

\noindent \textbf{4.2 Spatial and temporal synchrony: corridors of growth or gradients of diffusible cues?}

\noindent The mechanism of axonal pathfinding is commonly described by growth cones' ability to sense gradients of diffusible guidance cues in the environment. At the midline, a switch in growth cone responsiveness to gradients is explained by a change in growth cone surface receptor expression.  Earlier, though, axonal pathfinding was described mediated by substrate-bound molecules, which are a common guidance mechanism for migrating cells (reviewed and investigated by Akin and Zipursky, 2016; Batlle and Wilkinson, 2012; Bernhardt et al., 1992; Bovolenta and Dodd, 1990; Brose et al., 1999; Ch\'{e}dotal, 2019; de Ramon Franc\`{a}s et al., 2017; Kidd et al., 1999; Kolodkin and Tessier-Lavigne, 2011; Lentz et al., 1997; Mai et al., 2009; Moore et al., 2012; Stoeckli and Landmesser, 1995; Tanaka and Kirschner, 1995; Tessier-Lavigne and Goodman, 1996).

In this study, the striking stereotyped spatio-temporal wiring suggests, without providing direct evidence, that growth cones in the zebrafish spinal cord find their path by following predetermined substrate-bound tracks in the environment. Hence, though disregarding that there perhaps occur corrections of guidance errors of a smaller magnitude than the here defined 10 \textgreek{m}m type (Hutson and Chien, 2002), the very fast stereotyped and error-free wiring in many segments would from a mechanistic point of view seem easiest, or simplest, to accomplish if the growth cones were following corridors of substrate-bound cues (i.e. short-range cues or physically attached growth cues) rather than gradients of diffusible cues (i.e. long-range cues). For example, the three different stereotyped tracks of a) the obliquely ascending axons joining the DLF (Movie 2), b) the canonical commissurals crossing perpendicular to the midline (Movie 1 and 3) and c) all the ipsilateral axons, would otherwise require three different intermingled gradients to be set up by yet other cells that would need to be present at precise spatial locations to generate the three different spatial gradients of guidance cues. Additionally, erroneous midline crossings, corrected subsequently to ipsilateral growth, were never observed in the spinal cord. This indicates an underlying error-free process that likely could be accomplished by stereotyped corridors of growth. This discussion, on stereotypical guidance, concerns the mechanism by which it is achieved, since it unarguably is occurring \textit{in vivo}. Indeed, zebrafish resemble each other generation after generation, which is proof of a stereotypical process, in some cases even down to details observed at the microscopic level, as recently reported for the processes leading to the emergence of patterned neuronal population activity in the developing zebrafish spinal cord (Wan et al., 2019).

Recent and older data from other investigators also point to the importance of substrate-bound guidance mechanism. Substrates (e.g. NrCAM, laminin) are known to have a dramatic impact on axonal growth (Andersen, 2001; Lentz et al., 1997; Stoeckli and Landmesser, 1995; Stoeckli et al., 1997; Tanaka and Kirschner, 1995). In the mammalian forebrain, commissural midline crossing requires a transient population of midline glial cells (reviewed in (Comer et al., 2015). Akin and Zipursky (2016) reported for the wiring of the \textit{Drosophila} R8 photoreceptor cells that while a gradient of substrate-bound netrin was observed, netrin-DCC was not required for R8 chemoattraction but was found to be required for the R8 axon to remain attached to its final target. Both Slit and netrin are possibly \textit{in vivo} presented to the axons in a co-localized substrate-bound form rather than as freely diffusible ligands (Akin and Zipursky, 2016; Brankatschk and Dickson, 2006; Brose et al., 1999; Kidd et al., 1999; Zelina et al., 2014). Indeed, physical attachment of netrin to a substrate seems important for its ability to induce an attractive response via the DCC receptor (Endo et al., 2016; Moore et al., 2012). 

Hence, whether growth cones are explained to be guided by gradients of diffusible cues (i.e. long-range cues) or by substrate-bound corridors of growth (i.e. short-range cues) may in part be a question of semantics; likely of minor importance for the cellular mechanisms of axon guidance \textit{in vivo} (also discussed by Akin and Zipursky, 2016), but probably of importance for setting up and organizing axon guidance at an organismal scale in a 3D \textit{in vivo} environment. Thus, the various ligands may well be substrate-bound \textit{in vivo} (i.e. are membrane-bound, reviewed by Stoeckli and Landmesser, 1995). However, gradients of ligands are still involved \textit{in vivo}, and \textit{in vitro} experiments have demonstrated clearly that gradients of diffusible cues can guide axons \textit{in vitro}. Nevertheless, \textit{in vitro} experiments represent non-physiological conditions that may not properly mimic and cannot substitute for a 3D \textit{in vivo} environment. (reviewed in Aberle, 2019; de Ramon Franc\`{a}s et al., 2017; Endo et al., 2016; Kolodkin and Tessier-Lavigne, 2011).

\vspace{\baselineskip}

\noindent \textbf{4.3 Filopodia interactions with surroundings and axonal shaft bends: growth under tension}

The commissural axons avoid each other on the floor plate. The recordings suggest the avoidance is mediated by the growth cones' filopodia, which seem to serve as a steering device that appears to act repulsively by avoiding prolonged interactions when colliding with neighboring axonal shafts in the vicinity. However, upon reaching the DLF at the contralateral side, the axons will eventually fasciculate. What mediates this shift from avoidance to fasciculation upon floor plate crossing remains unknown, though similar observations were reported earlier (Bak and Fraser, 2003; Drapeau et al., 2002; Michalski et al., 2013; Moon and Gomez, 2005; Myers and Bastiani, 1993a;b). Axonal fasciculation and peripheral synapse formation likely could be analyzed further \textit{in vivo} with this nanometer-resolution light-sheet approach.

The sinusoidal-shaped shaft-bendings illustrate how wobbly the entire network is at this stage of development. The movies strongly suggest a direct role for filopodia in inducing these reversible bends. A previous study using fixed \textit{Xenopus} spinal cords also reported commissural shaft bends and proposed similarly that they are caused by filopodia interactions (Mason and Wang, 1997; Moon and Gomez, 2005). 

A second shaft-bending mechanism appears likely, based on some of the recordings. Notably at guidance points, namely, for example when the axons reach the contralateral side, the growth cones sometimes stall as if loosing traction and the trailing shaft is then observed to bend (e.g. Movie 4 label N2 and N3). The shafts of growing axons have been reported to be under pulling tension by being stretched out like an elastic band by advancing growth cones. Hence, it seems possible that the observed bends to some extent could be caused also by the advancing growth cones losing traction temporarily, thereby releasing the pulling tension on the shaft that would then bend like a relaxed elastic band (Athamneh et al., 2015; Athamneh and Suter, 2015; Holland et al., 2015; Mota and Herculano-Houzel, 2012; O'Toole et al., 2008). 

In contrast, the ipsilateral axons were not seen making sinusoidal-shaped bends in the available recordings. Their fast growth rates increase pulling tension on the shafts, which may change their flexural rigidity as compared with that of the commissural shafts; also, the ipsilateral growth cones did not seem to stall at any point. Perhaps because they are growing in a substrate-bound corridor of growth and have less guidance information to interpret as compared to commissural axons crossing the floor plate. Moreover, ipsilateral axons appeared in the movies to have few neighboring growth cones with filopodia that could collide with them.

It appears likely that the observed bends of the commissural shafts may be caused both by temporary loss of growth cone traction and by random collisions of filopodia from neighboring growth cones.

\vspace{\baselineskip}

\noindent \textbf{4.4 \textit{In vivo} observations of axonal growth under tension: support of tension-based cortical folding models?}

\noindent The observations here indicate axons \textit{in vivo} are growing under tension, as discussed in the preceding section. Axonal growth under tension \textit{in vivo} is of relevance to tension-based models of cortical folding that remain based on theoretical assumptions owing to the lack of experimental data and good model systems to investigate cortical morphogenesis. The mechanism(s) underlying cortical folding -- the formation of the gyri and sulci observed on the cerebral surface - remains under investigation though many different models and theories have been proposed. One tension-based theory of cortical morphogenesis posits that the bundles of axons in white matter, which are interconnecting the different brain areas, globally by tension 1) minimize the aggregate length of connections in the highly interconnected network of the brain (like in an optimized microchip circuit) and 2) counterbalance the hydrostatic pressure caused by the fluid produced by the ependymal cells that line the lumen of the neural canal (and later the choroid plexus in the ventricles), and thusly that radial tension along the parallel axonal fibers provide the forces that pull connected cortical regions closer together and thereby cause the observed folding of the cerebral cortex (Van Essen, 1997). Moreover, radial tension may be permissive for the enormous tangential and space requiring early cellular proliferation observed in the gray matter, intensively investigated and reviewed in recent decades (reviewed by Llinares-Benadero and Borrell (2019) and Pinson and Huttner (2021)), which in surface area exceeds the surface area of the underlying white matter and that shows species-dependent variability in thickness (Mota and Herculano-Houzel, 2015; Richman et al., 1975; Van Essen, 2020). In another brain folding tension modelling approach (Holland et al., 2015), aiming at a unified theory for gyrogenesis, it was investigated how the simulated brain would fold using experimental data from over-stretched axons \textit{in vitro} combined with information about the orientation of axons in the brain (known from diffusion spectrum imaging of the brain). Folding-patterns akin to cortical folding were obtained. Yet, the data suggested the stretched growing axons rapidly assumed a new equilibrium - they adapted quickly - when subject to stretch/tension, and as such that it rather may be the gray matter that is pulling on the white matter (Holland et al., 2015; Mota and Herculano-Houzel, 2012). However, as pointed out, the experimental data on stretch-induced axonal growth were from \textit{in vitro} experiments and the axons may behave differently \textit{in vivo}. \textit{In vivo }light-sheet experiments may be used to obtain better raw experimental \textit{in vivo} axonal growth and tension data parameters for use in such folding models and theoretical simulations. A need for \textit{in vivo} experiments to progress further the theoretical physical modelling of cortical folding morphogenesis is required, as the calculated output from theoretical simulation studies cannot be better than the real experimental \textit{in vivo} parameters used in the calculations, and the extent of tension as a primary mechanism of cortical folding remains an open question.

\vspace{\baselineskip}

\noindent \textbf{4.5 Uncovering the mechanisms of neural network formation}

\noindent An understanding of the many functions and facets of neuronal networks \textit{in vivo} requires an investigation of the mechanisms of network formation and a physical mapping of the networks. However, current understanding remains to a significant extent based on \textit{in vitro }data, and it is well known that \textit{in vitro }experiments may not reflect \textit{in vivo} processes, only being a projection of what could, and actually often does, occur \textit{in vivo}. (Andersen, 2005; Denk et al., 2012; Goodhill, 2016; Hildebrand et al., 2017; Mai et al., 2009; Nicol et al., 2011; Rosoff et al., 2004; Yuan et al., 2003).

\noindent The live recordings obtained in this study are \textit{in vivo }observations that are not prone to \textit{in vitro }artefacts, although transgenic FP-lines should be applied with caution as they are not completely equivalent to their wild-type counterparts (Lipt\'{a}k et al., 2019). By applying the transparent zebrafish in combination with the light-sheet platform, cell-type specific gene modification and single cell analysis, \textit{in vivo }analyses of neuronal wiring are within experimental reach.

\vspace{\baselineskip}
\vspace{\baselineskip}

\begin{small}
\noindent \textbf{Author's contribution}

\noindent S. Andersen performed and analyzed all experiments and data, conceived, wrote and published this study from 2015 to 2021.

\vspace{\baselineskip}

\noindent \textbf{Acknowledgements}

\noindent All experiments in this paper were performed from 2015 to 2016 at the SciLifeLab Biovis Facility. The SciLifeLab Zebrafish Facility and acquisition and maintenance of the light sheet microscopy used herein were made possible at UU by the Swedish Science for Life Laboratory initiative (https://www.SciLifeLab.se/). Thanks to the SciLifeLab BioVis staff for their helpful technical advice and equipment maintenance. Thanks to the SciLifeLab Zebrafish Technology Facility staff for caretaking of the zebrafish. Thanks to S.-I. Higashijima who in 2014 had kindly supplied UU with the transgenic dmrt3-gfp zebrafish used in this study. During the 12 months September 2015 to September 2016 stay at UU, S. Andersen was a non-employee supported for 12 months by a Swedish Beijerstiftelsen stipend administered by UU. 
\noindent Thanks to R. Francis, J. Jonsson, S. Edwards, 0. Fika, G. do Nacimiento, B. Aresh, P. Verveer, M. Hilscher, G. Alavioon, F. Freitag, C. Nagaraja, C. Restrepo, K. E. Wennstr\"{o}m, L. Shen and G. Khodus for help and discussions. Thanks to editors and reviewers at several journals for insights and constructive comments and to those who saw this research through to publication.

\vspace{\baselineskip}

\noindent \textbf{Conflict of interest statement}

\noindent The author declares no competing financial interests. A different version of this manuscript was published on-line in the \textit{Journal of Comparative Neurology} on August 30, 2017 and withdrawn on December 18, 2018, DOI: 10.1002/cne.24314. The content herein was presented at the AXON2017 Molecular and Cellular Mechanism of Neural Circuit Assembly meeting September 11-14 2017. A different version of this manuscript was published on-line in \textit{Journal of Integrative Neuroscience} on December 30, 2019 and retracted on September 30, 2020, DOI: 10.31083/j.jin.20\\19.04.1191. UU is herein mentioned to state the locale where this research was carried out from 2015 to 2016 but this does not imply an endorsement by UU, and UU explicitly does not endorse S. Andersen or this paper, solely to indicate these facts.\end{small}

\eject

\noindent \textbf{REFERENCES}

\begin{hangparas}{1.27cm}{1}

Aberle, H. (2019) Axon Guidance and Collective Cell Migration by Substrate-Derived Attractants. Front Mol Neurosci 12, 148. https://doi.org/10.3389/fnmol.2019.00148

Akin, O. and Zipursky, S. L. (2016) Frazzled Promotes Growth Cone Attachment at the Source of a Netrin Gradient in the Drosophila Visual
System. Elife 5, e20762. https://doi.org/10.7554/eLife.20762

Andersen, E. F. and Halloran, M. C. (2012) Centrosome Movements in Vivo Correlate with Specific Neurite Formation Downstream of Lim Homeodomain Transcription Factor Activity. Development 139, 3590-3599.\\https://doi.org/10.1242/dev.081513

Andersen, S. S. (2001) Preparation of Dissociated Zebrafish Spinal Neuron Cultures. Methods in Cell Science 23, 205-209. https://doi.org/10.1023/a:1016349232389

Andersen, S. S. (2003) Model System. Methods Cell Sci vii-viii.\\https://doi.org/10.1023/b:mics.0000006950.18528.c5

Andersen, S. S. (2005) The Search and Prime Hypothesis for Growth Cone Turning. BioEssays 27, 86-90. https://doi.org/10.1002/bies.20154

Andersen, S. S. and Bi, G. Q. (2000) Axon Formation: A Molecular Model for the Generation of Neuronal Polarity. Bioessays 22, 172-179.\\https://doi.org/10.1002/(sici)1521-1878(200002)22:2<172::Aid-bies8>3.0.Co;2-q

Andersen, S. S., Buendia, B., Dom\'{i}nguez, J. E., Sawyer, A. and Karsenti, E. (1994) Effect on Microtubule Dynamics of Xmap230, a Microtubule-Associated Protein Present in Xenopus Laevis Eggs and Dividing Cells. J Cell Biol 127, 1289-1299.\\https://doi.org/10.1083/jcb.127.5.1289

Athamneh, A. I., Cartagena-Rivera, A. X., Raman, A. and Suter, D. M. (2015) Substrate Deformation Predicts Neuronal Growth Cone Advance. Biophys J 109, 1358-1371. https://doi.org/10.1016/j.bpj.2015.08.013

Athamneh, A. I. and Suter, D. M. (2015) Quantifying Mechanical Force in Axonal Growth and Guidance. Front Cell Neurosci 9, 359.\\https://doi.org/10.3389/fncel.2015.00359

Bak, M. and Fraser, S. E. (2003) Axon Fasciculation and Differences in Midline Kinetics between Pioneer and Follower Axons within Commissural Fascicles. Development 130, 4999-5008. https://doi.org/10.1242/dev.00713

Batlle, E. and Wilkinson, D. G. (2012) Molecular Mechanisms of Cell Segregation and Boundary Formation in Development and Tumorigenesis. Cold Spring Harbor Perspectives in Biology 4, a008227. https://doi.org/10.1101/cshperspect.a008227

Bernhardt, R. R. (1994) Ipsi- and Contralateral Commissural Growth Cones React Differently to the Cellular Environment of the Ventral Zebrafish Spinal Cord. Journal of Comparative Neurology 350, 122-132. https://doi.org/10.1002/cne.903500109

Bernhardt, R. R., Chitnis, A. B., Lindamer, L. and Kuwada, J. Y. (1990) Identification of Spinal Neurons in the Embryonic and Larval Zebrafish. Journal of Comparative Neurology 302, 603-616. https://doi.org/10.1002/cne.903020315

Bernhardt, R. R., Nguyen, N. and Kuwada, J. Y. (1992) Growth Cone Guidance by Floor Plate Cells in the Spinal Cord of Zebrafish Embryos. Neuron 8, 869-882.\\https://doi.org/10.1016/0896-6273(92)90201-n

Bonner, J., Letko, M., Nikolaus, O. B., Krug, L., Cooper, A., Chadwick, B., Conklin, P., Lim, A., Chien, C. B. and Dorsky, R. I. (2012) Midline Crossing Is Not Required for Subsequent Pathfinding Decisions in Commissural Neurons. Neural Development 7, 18. https://doi.org/10.1186/1749-8104-7-18

Bovolenta, P. and Dodd, J. (1990) Guidance of Commissural Growth Cones at the Floor Plate in Embryonic Rat Spinal Cord. Development 109, 435-447.\\ https://doi.org/10.1242/dev.109.2.435

Brankatschk, M. and Dickson, B. J. (2006) Netrins Guide Drosophila Commissural Axons at Short Range. Nat Neurosci 9, 188-194. https://doi.org/10.1038/nn1625Bremer, J. and Granato, M. (2016) Myosin Phosphatase Fine-Tunes Zebrafish Motoneuron Position During Axonogenesis. PLoS Genetics 12, e1006440.\\https://doi.org/10.1371/journal.pgen.1006440

Brose, K., Bland, K. S., Wang, K. H., Arnott, D., Henzel, W., Goodman, C. S., Tessier-Lavigne, M. and Kidd, T. (1999) Slit Proteins Bind Robo Receptors and Have an Evolutionarily Conserved Role in Repulsive Axon Guidance. Cell 96, 795-806.https://doi.org/10.1016/s0092-8674(00)80590-5

Buckley, C. E., Ren, X., Ward, L. C., Girdler, G. C., Araya, C., Green, M. J., Clark, B. S., Link, B. A. and Clarke, J. D. (2013) Mirror-Symmetric Microtubule Assembly and Cell Interactions Drive Lumen Formation in the Zebrafish Neural Rod. EMBO Journal 32, 30-44. https://doi.org/10.1038/emboj.2012.305

Burgess, H. A., Johnson, S. L. and Granato, M. (2009) Unidirectional Startle Responses and Disrupted Left-Right Co-Ordination of Motor Behaviors in Robo3 Mutant Zebrafish. Genes, Brain and Behavior 8, 500-511. https://doi.org/10.1111/j.1601-183X.2009.00499.x

Cearns, M. D., Escuin, S., Alexandre, P., Greene, N. D. and Copp, A. J. (2016) Microtubules, Polarity and Vertebrate Neural Tube Morphogenesis. J Anat 229, 63-74.\\https://doi.org/10.1111/joa.12468

Challa, A. K., McWhorter, M. L., Wang, C., Seeger, M. A. and Beattie, C. E. (2005) Robo3 Isoforms Have Distinct Roles During Zebrafish Development. Mech Dev 122, 1073-1086. https://doi.org/10.1016/j.mod.2005.06.006

Ch\'{e}dotal, A. (2019) Roles of Axon Guidance Molecules in Neuronal Wiring in the Developing Spinal Cord. Nature Reviews Neuroscience 20, 380-396.\\ https://doi.org/10.1038/s41583-019-0168-7

Comer, J. D., Alvarez, S., Butler, S. J. and Kaltschmidt, J. A. (2019) Commissural Axon Guidance in the Developing Spinal Cord: From Cajal to the Present Day. Neural Development 14, 1-16. https://doi.org/10.1186/s13064-019-0133-1

Comer, J. D., Pan, F. C., Willet, S. G., Haldipur, P., Milen, K. J., Wright, C. V. E. and Kaltschmidt, J. A. (2015) Sensory and Spinal Inhibitory Dorsal Midline Crossing Is Independent of Robo3. Frontiers in Neural Circuits 9, 36.\\https://doi.org/10.3389/fncir.2015.00036

de Ramon Franc\`{a}s, G., Zu\~{n}iga, N. R. and Stoeckli, E. T. (2017) The Spinal Cord Shows the Way - How Axons Navigate Intermediate Targets. Dev Biol 432, 43-52.\\https://doi.org/10.1016/j.ydbio.2016.12.002

de Vito, G., Turrini, L., M\"{u}llenbroich, C., Ricci, P., Sancataldo, G., Mazzamuto, G., Tiso, N., Sacconi, L., Fanelli, D., Silvestri, L., Vanzi, F. and Pavone, F. S. (2021) Fast Whole-Brain Imaging of Seizures in       Zebrafish Larvae by Two-Photon Light Sheet Miroscopy. arXiv:201209660v3 [q-bioQM]. https://arxiv.org/abs/2012.09660v3

Denk, W., Briggman, K. L. and Helmstaedter, M. (2012) Structural Neurobiology: Missing Link to a Mechanistic Understanding of Neural Computation. Nature Reviews Neuroscience 13, 351-358. https://doi.org/10.1038/nrn3169

Drapeau, P., Saint-Amant, L., Buss, R. R., Chong, M., McDearmid, J. R. and Brustein, E. (2002) Development of the Locomotor Network in Zebrafish. Prog Neurobiol 68, 85-111. https://doi.org/10.1016/s0301-0082(02)00075-8

Dumoulin, A., Zu\~{n}iga, N. R. and Stoeckli, E. T. (2021) Axon Guidance at the Spinal Cord Midline-a Live Imaging Perspective. J Comp Neurol 529, 2517-2538.\\https://doi.org/10.1002/cne.25107

Endo, M., Hattori, M., Toriyabe, H., Ohno, H., Kamiguchi, H., Iino, Y. and Ozawa, T. (2016) Optogenetic Activation of Axon Guidance Receptors Controls Direction of Neurite Outgrowth. Sci Rep 6, 23976. https://doi.org/10.1038/srep23976

Erdogan, B., Bearce, E. A. and Lowery, L. A. (2021) Live Imaging of Cytoskeletal Dynamics in Embryonic Xenopus Laevis Growth Cones and Neural Crest Cells. Cold Spring Harb Protoc 2021, pdb prot104463.\\https://doi.org/10.1101/pdb.prot104463

Erdogan, B., Ebbert, P. T. and Lowery, L. A. (2016) Using Xenopus Laevis Retinal and Spinal Neurons to Study Mechanisms of Axon Guidance in Vivo and in Vitro. Semin Cell Dev Biol 51, 64-72. https://doi.org/10.1016/j.semcdb.2016.02.003

Fischer, R. S., Lam, P. Y., Huttenlocher, A. and Waterman, C. M. (2019) Filopodia and Focal Adhesions: An Integrated System Driving Branching Morphogenesis in Neuronal Pathfinding and Angiogenesis. Dev Biol 451, 86-95.\\https://doi.org/10.1016/j.ydbio.2018.08.015

Gaynes, J. A., Otsuna, H., Campbell, D. S., Manfredi, J. P., Levine, E. M. and Chien, C. B. (2015) The Rna Binding Protein Igf2bp1 Is Required for Zebrafish Rgc Axon Outgrowth in Vivo. PLoS One 10, e0134751.\\https://doi.org/10.1371/journal.pone.0134751

Geldmacher-Voss, B., Reugels, A. M., Pauls, S. and Campos-Ortega, J. A. (2003) A 90-Degree Rotation of the Mitotic Spindle Changes the Orientation of Mitoses of Zebrafish Neuroepithelial Cells. Development 130, 3767-3780.\\https://doi.org/10.1242/dev.00603

Goodhill, G. J. (2016) Can Molecular Gradients Wire the Brain? Trends in Neurosciences 39, 202-211. https://doi.org/10.1016/j.tins.2016.01.009

Griener, A., Zhang, W., Kao, H., Haque, F. and Gosgnach, S. (2017) Anatomical and Electrophysiological Characterization of a Population of Di6 Interneurons in the Neonatal Mouse Spinal Cord. Neuroscience 362, 47-59.\\https://doi.org/10.1016/j.neuroscience.2017.08.031

Hale, M. E., Ritter, D. A. and Fetcho, J. R. (2001) A Confocal Study of Spinal Interneurons in Living Larval Zebrafish. Journal of Comparative Neurology 437, 1-16. https://doi.org/10.1002/cne.1266

Higashijima, S., Schaefer, M. and Fetcho, J. R. (2004) Neurotransmitter Properties of Spinal Interneurons in Embryonic and Larval Zebrafish. Journal of Comparative Neurology 480, 19-37. https://doi.org/10.1002/cne.20279

Hildebrand, D. G. C., Cicconet, M., Torres, R. M., Choi, W., Quan, T. M., Moon, J., Wetzel, A. W., Scott Champion, A., Graham, B. J., Randlett, O., Plummer, G. S., Portugues, R., Bianco, I. H., Saalfeld, S., Baden, A. D., Lillaney, K., Burns, R., Vogelstein, J. T., Schier, A. F., Lee, W. A., Jeong, W. K., Lichtman, J. W. and Engert, F. (2017) Whole-Brain Serial-Section Electron Microscopy in Larval Zebrafish. Nature 545, 345-349. https://doi.org/10.1038/nature22356

Holland, M. A., Miller, K. E. and Kuhl, E. (2015) Emerging Brain Morphologies from Axonal Elongation. Annals of Biomedical Engineering 43, 1640-1653.\\https://doi.org/10.1007/s10439-015-1312-9

Hutson, L. D. and Chien, C. B. (2002) Pathfinding and Error Correction by Retinal Axons: The Role of Astray/Robo2. Neuron 33, 205-217. https://doi.org/10.1016/s0896-6273(01)00579-7

Jaworski, A., Tom, I., Tong, R. K., Gildea, H. K., Koch, A. W., Gonzalez, L. C. and Tessier-Lavigne, M. (2015) Operational Redundancy in Axon Guidance through the Multifunctional Receptor Robo3 and Its Ligand Nell2. Science 350, 961-965.\\
https://doi.org/10.1126/science.aad2615

Kastenhuber, E., Kern, U., Bonkowsky, J. L., Chien, C. B., Driever, W. and Schweitzer, J. (2009) Netrin-Dcc, Robo-Slit, and Heparan Sulfate Proteoglycans Coordinate Lateral Positioning of Longitudinal Dopaminergic Diencephalospinal Axons. The Journal of Neuroscience 29, 8914-8926. https://doi.org/10.1523/jneurosci.0568-09.2009

Kaufmann, A., Mickoleit, M., Weber, M. and Huisken, J. (2012) Multilayer Mounting Enables Long-Term Imaging of Zebrafish Development in a Light Sheet Microscope. Development 139, 3242-3247. https://doi.org/10.1242/dev.082586

Keller, P. J. and Ahrens, M. B. (2015) Visualizing Whole-Brain Activity and Development at the Single-Cell Level Using Light-Sheet Microscopy. Neuron 85, 462-483.\\
https://doi.org/10.1016/j.neuron.2014.12.039

Keller, P. J., Schmidt, A. D., Wittbrodt, J. and Stelzer, E. H. (2008) Reconstruction of Zebrafish Early Embryonic Development by Scanned Light Sheet Microscopy. Science 322, 1065-1069. https://doi.org/10.1126/science.1162493

Key, B. and Devine, C. A. (2003) Zebrafish as an Experimental Model: Strategies for Developmental and Molecular Neurobiology Studies. Methods in Cell Science 25, 1-6. https://doi.org/10.1023/b:mics.0000006849.98007.03

Kidd, T., Bland, K. S. and Goodman, C. S. (1999) Slit Is the Midline Repellent for the Robo Receptor in Drosophila. Cell 96, 785-794. https://doi.org/10.1016/s0092-8674(00)80589-9

Kim, M., Farmer, W. T., Bjorke, B., McMahon, S. A., Fabre, P. J., Charron, F. and Mastick, G. S. (2014) Pioneer Midbrain Longitudinal Axons Navigate Using a Balance of Netrin Attraction and Slit Repulsion. Neural Development 9, 17.\\https://doi.org/10.1186/1749-8104-9-17

Kimmel, C. B., Ballard, W. W., Kimmel, S. R., Ullmann, B. and Schilling, T. F. (1995) Stages of Embryonic Development of the Zebrafish. Developmental Dynamics 203, 253-310. https://doi.org/10.1002/aja.1002030302

Kimmel, C. B., Warga, R. M. and Kane, D. A. (1994) Cell Cycles and Clonal Strings During Formation of the Zebrafish Central Nervous System. Development 120, 265-276.https://doi.org/10.1242/dev.120.2.265

Kishore, S., Cadoff, E. B., Agha, M. A. and McLean, D. L. (2020) Orderly Compartmental Mapping of Premotor Inhibition in the Developing Zebrafish Spinal Cord. Science 370, 431-436. https://doi.org/10.1126/science.abb4608

Kolodkin, A. L. and Tessier-Lavigne, M. (2011) Mechanisms and Molecules of Neuronal Wiring: A Primer. Cold Spring Harbor Perspectives in Biology 3, \\
https://doi.org/10.1101/cshperspect.a001727

Kuwada, J. Y., Bernhardt, R. R. and Chitnis, A. B. (1990a) Pathfinding by Identified Growth Cones in the Spinal Cord of Zebrafish Embryos. The Journal of Neuroscience 10, 1299-1308. https://doi.org/10.1523/JNEUROSCI.10-04-01299.1990

Kuwada, J. Y., Bernhardt, R. R. and Nguyen, N. (1990b) Development of Spinal Neurons and Tracts in the Zebrafish Embryo. Journal of Comparative Neurology 302, 617-628. https://doi.org/10.1002/cne.903020316

Lee, T. J., Lee, J. W., Haynes, E. M., Eliceiri, K. W. and Halloran, M. C. (2017) The Kinesin Adaptor Calsyntenin-1 Organizes Microtubule Polarity and Regulates Dynamics During Sensory Axon Arbor Development. Front Cell Neurosci 11, 107.\\
https://doi.org/10.3389/fncel.2017.00107

Lentz, S. I., Miner, J. H., Sanes, J. R. and Snider, W. D. (1997) Distribution of the Ten Known Laminin Chains in the Pathways and Targets of Developing Sensory Axons. Journal of Comparative Neurology 378, 547-561.\\ https://doi.org/10.1002/(sici)1096-9861(19970224)378:4<547::aid-cne9>3.0.co;2-2

Lewis, K. E. and Eisen, J. S. (2003) From Cells to Circuits: Development of the Zebrafish Spinal Cord. Prog Neurobiol 69, 419-449.\\https://doi.org/10.1016/s0301-0082(03)00052-2

Li, Z., Ptak, D., Zhang, L., Walls, E. K., Zhong, W. and Leung, Y. F. (2012) Phenylthiourea Specifically Reduces Zebrafish Eye Size. PLoS One 7, e40132.\\
https://doi.org/10.1371/journal.pone.0040132

Lipt\'{a}k, N., B\H{o}sze, Z. and Hiripi, L. (2019) Gfp Transgenic Animals in Biomedical Research: A Review of Potential Disadvantages. Physiological Research 68, 525-530.\\
https://doi.org/10.33549/physiolres.934227

Liu, Y. and Halloran, M. C. (2005) Central and Peripheral Axon Branches from One Neuron Are Guided Differentially by Semaphorin3d and Transient Axonal Glycoprotein-1. The Journal of Neuroscience 25, 10556-10563.\\https://doi.org/10.1523/jneurosci.2710-05.2005

Llinares-Benadero, C. and Borrell, V. (2019) Deconstructing Cortical Folding: Genetic, Cellular and Mechanical Determinants. Nat Rev Neurosci 20, 161-176.\\https://doi.org/10.1038/s41583-018-0112-2

Mai, J., Fok, L., Gao, H., Zhang, X. and Poo, M. M. (2009) Axon Initiation and Growth Cone Turning on Bound Protein Gradients. The Journal of Neuroscience 29, 7450-7458. https://doi.org/10.1523/jneurosci.1121-09.2009

Mason, C. A. and Wang, L.-C. (1997) Growth Cone Form Is Behavior-Specific and, Consequently, Position-Specific Along the Retinal Axon Pathway. The Journal of Neuroscience 17, 1086-1100. https://doi.org/10.1523/JNEUROSCI.17-03-01086.1997

Michalski, N., Babai, N., Renier, N., Perkel, D. J., Chedotal, A. and Schneggenburger, R. (2013) Robo3-Driven Axon Midline Crossing Conditions Functional Maturation of a Large Commissural Synapse. Neuron 78, 855-868.\\ https://doi.org/10.1016/j.neuron.2013.04.006

Moon, M.-s. and Gomez, T. M. (2005) Adjacent Pioneer Commissural Interneuron Growth Cones Switch from Contact Avoidance to Axon Fasciculation after Midline Crossing. Dev Biol 228, 474-486. https://doi.org/10.1016/j.ydbio.2005.09.049

Moore, S. W., Zhang, X., Lynch, C. D. and Sheetz, M. P. (2012) Netrin-1 Attracts Axons through Fak-Dependent Mechanotransduction. The Journal of Neuroscience 32, 11574-11585. https://doi.org/10.1523/jneurosci.0999-12.2012

Mota, B. and Herculano-Houzel, S. (2012) How the Cortex Gets Its Folds: An inside-out, Connectivity-Driven Model for the Scaling of Mammalian Cortical Folding. Frontiers in Neuroanatomy 6, 13. https://doi.org/10.3389/fnana.2012.00003

Mota, B. and Herculano-Houzel, S. (2015) Brain Structure. Cortical Folding Scales Universally with Surface Area and Thickness, Not Number of Neurons. Science 349, 74-77. https://doi.org/10.1126/science.aaa9101

Murray, M. J. and Whitington, P. M. (1999) Effects of Roundabout on Growth Cone Dynamics, Filopodial Length, and Growth Cone Morphology at the Midline and Throughout the Neuropile. The Journal of Neuroscience 19, 7901-7912.\\ https://doi.org/10.1523/JNEUROSCI.19-18-07901.1999

Myers, P. Z. and Bastiani, M. J. (1993a) Cell-Cell Interactions During the Migration of an Identified Commissural Growth Cone in the Embryonic Grasshopper. The Journal of Neuroscience 13, 115-126. https://doi.org/10.1523/JNEUROSCI.13-01-00115.1993

Myers, P. Z. and Bastiani, M. J. (1993b) Growth Cone Dynamics During the Migration of an Identified Commissural Growth Cone. The Journal of Neuroscience 13, 127-143. https://doi.org/10.1523/JNEUROSCI.13-01-00127.1993

Nagel, A. N., Marshak, S., Manitt, C., Santos, R. A., Piercy, M. A., Mortero, S. D., Shirkey-Son, N. J. and Cohen-Cory, S. (2015) Netrin-1 Directs Dendritic Growth and Connectivity of Vertebrate Central Neurons in Vivo. Neural Development 10, 14.\\https://doi.org/10.1186/s13064-015-0041-y

Nicol, X., Hong, K. P. and Spitzer, N. C. (2011) Spatial and Temporal Second Messenger Codes for Growth Cone Turning. Proceedings of the National Academy of Sciences of the United States of America 108, 13776-13781.\\https://doi.org/10.1073/pnas.1100247108

O'Toole, M., Lamoureux, P. and Miller, K. E. (2008) A Physical Model of Axonal Elongation: Force, Viscosity, and Adhesions Govern the Mode of Outgrowth. Biophys J 94, 2610-2620. https://doi.org/biophysj.107.117424

Paix\~{a}o, S., Balijepalli, A., Serradj, N., Niu, J., Luo, W., Martin, J. H. and Klein, R. (2013) Ephrinb3/Epha4-Mediated Guidance of Ascending and Descending Spinal Tracts. Neuron 80, 1407-1420. https://doi.org/10.1016/j.neuron.2013.10.006

Papan, C. and Campos-Ortega, J. A. (1994) On the Formation of the Neural Keel and Neural Tube in the Zebrafish Danio (Brachydanio) Rerio. Roux's Archive of Developmental Biology 203, 178-186. https://doi.org/10.1007/bf00636333

Pinson, A. and Huttner, W. B. (2021) Neocortex Expansion in Development and Evolution-from Genes to Progenitor Cell Biology. Curr Opin Cell Biol 73, 9-18.\\https://doi.org/10.1016/j.ceb.2021.04.008

Pittman, A. J., Law, M. Y. and Chien, C. B. (2008) Pathfinding in a Large Vertebrate Axon Tract: Isotypic Interactions Guide Retinotectal Axons at Multiple Choice Points. Development 135, 2865-2871. https://doi.org/10.1242/dev.025049

Plazas, P. V., Nicol, X. and Spitzer, N. C. (2013) Activity-Dependent Competition Regulates Motor Neuron Axon Pathfinding Via Plexina3. Proceedings of the National Academy of Sciences of the United States of America 110, 1524-1529.\\https://doi.org/10.1073/pnas.1213048110

Portera-Cailliau, C., Weimer, R. M., De Paola, V., Caroni, P. and Svoboda, K. (2005) Diverse Modes of Axon Elaboration in the Developing Neocortex. PLoS Biology 3, e272. https://doi.org/10.1371/journal.pbio.0030272

Qin, X., Chen, C., Wang, L., Chen, X., Liang, Y., Jin, X., Pan, W., Liu, Z., Li, H. and Yang, G. (2021) In-Vivo 3d Imaging of Zebrafish's Intersegmental Vessel Development by a Bi-Directional Light-Sheet Illumination Microscope. Biochem Biophys Res Commun 557, 8-13. https://doi.org/10.1016/j.bbrc.2021.03.160

Richman, D. P., Stewart, R. M., Hutchinson, J. W. and Caviness, V. S., Jr. (1975) Mechanical Model of Brain Convolutional Development. Science 189, 18-21.\\https://doi.org/10.1126/science.1135626

Rojo, F. P., Nyman, R. K. M., Johnson, A. A. T. J., Navarro, M. P., Ryan, M. H., Erskine, W. and Kaur, P. (2018) Crispr-Cas Systems: Ushering in the New Genome Editing Era. Bioengineerd 9, 214-221. https://doi.org/10.1080/21655979.2018.1470720

Rosoff, W. J., Urbach, J. S., Esrick, M. A., McAllister, R. G., Richards, L. J. and Goodhill, G. J. (2004) A New Chemotaxis Assay Shows the Extreme Sensitivity of Axons to Molecular Gradients. Nat Neurosci 7, 678-682. https://doi.org/10.1038/nn1259

Sakai, N. and Kaprielian, Z. (2012) Guidance of Longitudinally Projecting Axons in the Developing Central Nervous System. Front Mol Neurosci 5, 59.\\ https://doi.org/10.3389/fnmol.2012.00059

Satou, C., Sugioka, T., Uemura, Y., Shimazaki, T., Zmarz, P., Kimura, Y. and Higashijima, S. I. (2020) Functional Diversity of Glycinergic Commissural Inhibitory Neurons in Larval Zebrafish. Cell Rep 30, 3036-3050 e3034.\\https://doi.org/10.1016/j.celrep.2020.02.015

Satou, C., Kimura, Y., Hirata, H., Suster, M. L., Kawakami, K. and Higashijima, S. (2013) Transgenic Tools to Characterize Neuronal Properties of Discrete Populations of Zebrafish Neurons. Development 140, 3927-3931.\\https://doi.org/10.1242/dev.099531

Satou, C., Kimura, Y., Kohashi, T., Horikawa, K., Takeda, H., Oda, Y. and Higashijima, S. (2009) Functional Role of a Specialized Class of Spinal Commissural Inhibitory Neurons During Fast Escapes in Zebrafish. J Neurosci 29, 6780-6793.\\
https://doi.org/10.1523/jneurosci.0801-09.2009

Schulte-Merker, S. and Stainier, D. Y. R. (2014) Out with the Old, in with the New: Reassessing Morpholino Knockdowns in Light of Genome Editing Technology. Development 141, 3103-3104. https://doi.org/10.1242/dev.112003

Schweitzer, J., Lohr, H., Bonkowsky, J. L., Hubscher, K. and Driever, W. (2013) Sim1a and Arnt2 Contribute to Hypothalamo-Spinal Axon Guidance by Regulating Robo2 Activity Via a Robo3-Dependent Mechanism. Development 140, 93-106.\\
https://doi.org/10.1242/dev.087825

Shah, A. N., Davey, C. F., Whitebirch, A. C., Miller, A. C. and Moens, C. B. (2015) Rapid Reverse Genetic Screening Using Crispr in Zebrafish. Nat Methods 12, 535-540. https://doi.org/10.1038/nmeth.3360

Short, C. A., Onesto, M. M., Rempel, S. K., Catlett, T. S. and Gomez, T. M. (2021) Familiar Growth Factors Have Diverse Roles in Neural Network Assembly. Curr Opin Neurobiol 66, 233-239. https://doi.org/10.1016/j.conb.2020.12.016

Sretavan, D. W. and Reichardt, L. F. (1993) Time-Lapse Video Analysis of Retinal Ganglion Cell Axon Pathfinding at the Mammalian Optic Chiasm: Growth Cone Guidance Using Intrinsic Chiasm Cues. Neuron 10, 761-777.\\ https://doi.org/10.1016/0896-6273(93)90176-r

Staiger, E. A., Almen, M. S., Promerova, M., Brooks, S., Cothran, E. G., Imsland, F., Jaderkvist Fegraeus, K., Lindgren, G., Mehrabani Yeganeh, H., Mikko, S., Vega-Pla, J. L., Tozaki, T., Rubin, C. J. and Andersson, L. (2017) The Evolutionary History of the Dmrt3 'Gait Keeper' Haplotype. Animal Genetics 48, 551-559. https://doi.org/10.1111/age.12580

Stoeckli, E. T. (2018) Understanding Axon Guidance: Are We Nearly There Yet? Development 145, https://doi.org/10.1242/dev.151415

Stoeckli, E. T. and Landmesser, L. T. (1995) Axonin-1, Nr-Cam, and Ng-Cam Play Different Roles in the in Vivo Guidance of Chick Commissural Neurons. Neuron 14, 1165-1179. https://doi.org/10.1016/0896-6273(95)90264-3

Stoeckli, E. T., Sonderegger, P., Pollerberg, G. E. and Landmesser, L. T. (1997) Interference with Axonin-1 and Nrcam Interactions Unmasks a Floor-Plate Activity Inhibitory for Commissural Axons. Neuron 18, 209-221. https://doi.org/10.1016/s0896-6273(00)80262-7

Tanaka, E. and Kirschner, M. W. (1995) The Role of Microtubules in Growth Cone Turning at Substrate Boundaries. Journal of Cell Biology 128, 127-137.\\ https://doi.org/10.1083/jcb.128.1.127

Tessier-Lavigne, M. and Goodman, C. S. (1996) The Molecular Biology of Axon Guidance. Science 274, 1123-1133. https://doi.org/10.1126/science.274.5290.1123

Van Essen, D. C. (1997) A Tension-Based Theory of Morphogenesis and Compact Wiring in the Central Nervous System. Nature 385, 313-318.\\https://doi.org/10.1038/385313a0

Van Essen, D. C. (2020) A 2020 View of Tension-Based Cortical Morphogenesis. Proc Natl Acad Sci U S A 117, 32868-32879. https://doi.org/10.1073/pnas.2016830117

Wan, Y., Wei, Z., Looger, L. L., Koyama, M., Druckmann, S. and Keller, P. J. (2019) Single-Cell Reconstruction of Emerging Population Activity in an Entire Developing Circuit. Cell 179, 355-372. https://doi.org/10.1016/j.cell.2019.08.039

Weber, M., Mickoleit, M. and Huisken, J. (2014) Multilayer Mounting for Long-Term Light Sheet Microscopy of Zebrafish. Journal of Visual Experiments: JoVE Feb 27, e51119. https://doi.org/10.3791/51119

Weinger, J. G., Greenberg, M. L., Matheu, M. P., Parker, I., Walsh, C. M., Lane, T. E. and Cahalan, M. D. (2015) Two-Photon Imaging of Cellular Dynamics in the Mouse Spinal Cord. Journal of Visual Experiments: JoVE https://doi.org/10.3791/52580

Yaginuma, H., Homma, S., K\"{u}nzi, R. and Oppenheim, R. W. (1991) Pathfinding by Growth Cones of Commissural Interneurons in the Chick Embryo Spinal Cord: A Light and Electron Microscopic Study. J Comp Neurol 304, 78-102.\\
https://doi.org/10.1002/cne.903040107

Yuan, X. B., Jin, M., Xu, X., Song, Y. Q., Wu, C. P., Poo, M. M. and Duan, S. (2003) Signalling and Crosstalk of Rho Gtpases in Mediating Axon Guidance. Nat Cell Biol 5, 38-45. https://doi.org/10.1038/ncb895

Zeisel, A., Munoz-Manchado, A. B., Codeluppi, S., Lonnerberg, P., La Manno, G., Jureus, A., Marques, S., Munguba, H., He, L., Betsholtz, C., Rolny, C., Castelo-Branco, G., Hjerling-Leffler, J. and Linnarsson, S. (2015) Brain Structure. Cell Types in the Mouse Cortex and Hippocampus Revealed by Single-Cell Rna-Seq. Science 347, 1138-1142. https://doi.org/10.1126/science.aaa1934
Zelina, P., Blockus, H., Zagar, Y., Peres, A., Friocourt, F., Wu, Z., Rama, N., Fouquet, C., Hohenester, E., Tessier-Lavigne, M., Schweitzer, J., Roest Crollius, H. and Chedotal, A. (2014) Signaling Switch of the Axon Guidance Receptor Robo3 During Vertebrate Evolution. Neuron 84, 1258-1272. https://doi.org/10.1016/j.neuron.2014.11.004

\end{hangparas}

\eject

\noindent \textbf{Supporting Information}

\vspace{\baselineskip}

\noindent Graphical Abstract

\noindent \textbf{\includegraphics*[width=6.12in, height=8.00in]{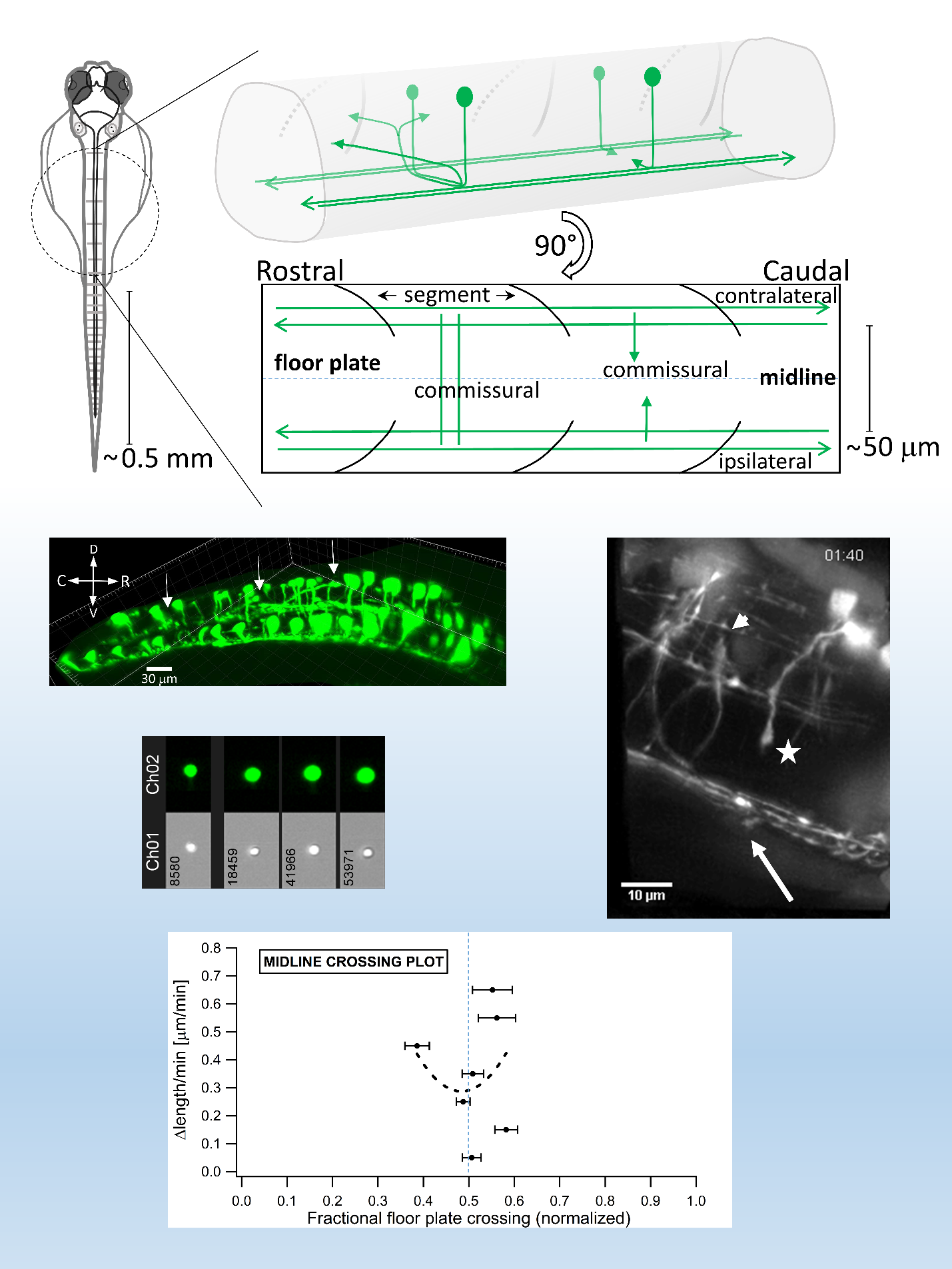}}

\noindent \begin{small} \textbf{Graphical Abstract Legend: }The midline represents a major guidance point for neurons during early wiring of the vertebrate nervous system. Light-sheet microscopy experiments show real time \textit{in vivo} recordings of midline crossing commissural axons in the zebrafish spinal cord, and indicate substrate bound corridors of preferential growth rather than diffusible guidance cues may predominate ipsilateral axonal guidance. \end{small} 

\vspace{\baselineskip}

\noindent \textbf{Movie S1}. \textbf{Developmental events just prior to midline crossing} \textbf{}

\noindent The events in this movie (Movie S1) are just prior in development to the outgrowth of the spinal cord axons shown in Movie 1. The recordings confirm previous accounts using antibody stained micrometer thick serial sections, widefield and confocal microscopy (Buckley et al., 2013; Geldmacher-Voss et al., 2003; Kimmel et al., 1994; Papan and Campos-Ortega, 1994), and demonstrate the versatility of the light-sheet microscopy, as a complement to confocal microscopy, also for studies of this kind. No cell divisions were observed during these recordings (Buckley et al., 2013); the divisions would be occurring at the apical side of the epithelium, wherefore the daughter cell would easily become located on the other side of the midline by division across the midline in a so-called C-division, for midline Crossing division (Buckley et al., 2013; Cearns et al., 2016; Kimmel et al., 1994; Papan and Campos-Ortega, 1994). The nascent neuronal cell bodies are seen moving and aligning bilaterally along the midline (Movie S1). The apical sides of the cells are facing the midline, forming a neuroepithelium on either side of the midline, each held together by adherence junctions between the cells. Starting rostrally/anteriorly (to the right), the cells are then observed retracting from the midline, displacing to more lateral levels and with the concomitant opening up of a lumen. Forming thereby the future neurocoel, an event that has been staged to begin at the 17-18 somite stage (Papan and Campos-Ortega, 1994). 

\vspace{\baselineskip} 

\noindent \begin{small} \textbf{Movie\_S1} Legend. Late stages of neurulation, apical alignment at the midline and neurocoel formation. Viewing the movies by scrolling back and forth frame by frame is most informative (possible e.g. with Windows' `Movies \& TV' application). In this sequence, the cells have lined up in the neural rod (the cells are long and extended at this stage) with their apical side facing each other at the midline. It is seen how the neurocoel forms (starting right) and extends rostral (right) towards caudal (left) while the zebrafish is growing. The very rostral part of the spinal cord is seen appearing to the left at the end of the sequence. Length 9 hours 12 minutes in real-time, the scale bar is 50 \textgreek{m}m: \end{small} 

\noindent \includegraphics*[width=1.30in, height=1.30in]{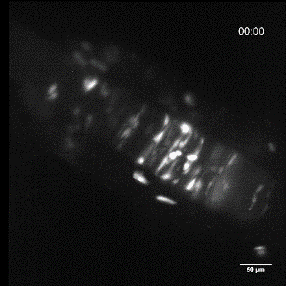}

\vspace{\baselineskip}

\noindent \begin{small} \textbf{Movie\_S2} Legend. In this Movie, the cell bodies have not been sliced away; showing axonal outgrowth from the cell body, crossing of the floor plate and contralateral growth. (viewing frame by frame is necessary to follow this description). Outgrowth of an axon (inside the oval window) from the still developing cell body (constant GFP expression, at constant laser power, causes the increased signal over time), extending downward and crossing the floor plate for then to grow up and continue at an oblique angle contralaterally. As the view is of a 3D event, subsequently MIP to 2D, it looks like the axon is making a sharp 180 degree turn whereas in fact it is not.  [Immediately upwards in the Movie of this neuron's cell body is a big bright cell body where a similar growth path can be discerned (in part blocked by its big cell body in this MIP)]. Contralaterally to this neuron, a cell body appears at t=2h, sending out a process downwards (ventrally); at t= 3h4min it then turns towards the floor plate (left in the Movie). At t = 4h, an ipsilateral longitudinal axon is seen shooting past this neuron's axon (underneath its cell body); it illustrates how crowded the neural network is, considering only GFP-labelled neurons are visible. Length 10 hours 32 minutes in real time, the bar is 10 \textgreek{m}m: \end{small} 

\noindent \includegraphics*[width=1.85in, height=1.30in]{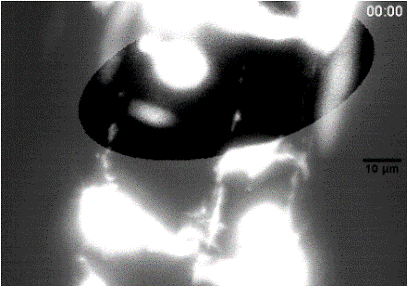}

\vspace{\baselineskip}

\noindent \begin{small} \textbf{Movie\_S3a} Legend. Ipsilateral guidance error correction in the future hindbrain region. At the beginning are seen cell movements in the future head, the zebrafish grows and towards the end of the sequence the cell bodies in the spinal cord are visible to the left (rostral is to the right). Numerous processes can be seen extending. At the time 10 hours and 40 minutes, ipsilateral axons are seen extending caudally in the top right (in the future hindbrain), most likely towards the medial longitudinal fasciculus (MLF) of the spinal cord. One of them extends too far and retracts upon reaching the midline at 14 hours for then to grow ipsilaterally, most likely towards the MLF of the spinal cord. A close view of that event is shown in Movie S3b. Notice the increase in the signal intensity over time due to the continuous expression of GFP during the recording (constant laser power). Length 16 hours 56 minutes in real time, the bar is 50 \textgreek{m}m: \end{small} 

\noindent \includegraphics*[width=1.56in, height=1.30in]{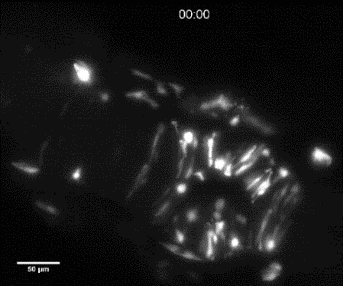}

\vspace{\baselineskip}

\noindent \begin{small} \textbf{Movie\_S3b} Legend. Zoom in of Movie S3a, showing an ipsilateral guidance error correction. A close view of the guidance error correction event described at 10 hours 40 minutes in Movie S3a. This axon withdraws rapidly back from the midline; then continues the correct ipsilateral path, most likely towards the MLF of the spinal cord, by inducing a new branch growing in the correct direction (well before withdrawal has fully completed). Length 2 hours 40 minutes in real time, the bar is 10 \textgreek{m}m: \end{small} 

\noindent \includegraphics*[width=1.10in, height=1.30in]{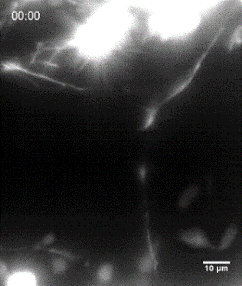}

\vspace{\baselineskip}

\noindent \begin{small} \textbf{Movie\_S4} Legend. Movie 4 without labelling: High resolution deconvolved sequence shows filopodia dynamics and growth paths of three commissural and one ipsilateral axons (Figure 3).\end{small} 

\noindent \begin{small} \textbf{Movie\_S5} Legend. Movie 5 without labelling: Commissural axonal shafts bend and wobble on the floor plate (Figure 4). \end{small} 

\noindent \begin{small} \textbf{Movie\_S6} Legend. Movie 6 without labelling: Commissural neurite shafts respond with sinusoidal-shaped bends upon contact by filopodia tips (Figure 4).\end{small} 

\vspace{\baselineskip} 

\noindent \begin{small} \textbf{Movie\_S7} Legend. Schematic showing how the different layers of the spinal cord may be uncovered and studied by slicing the dorsal cell bodies away, zooming in and adjusting intensity, the bar in the left corner varies with the zoom ratio:\end{small} 

\noindent \includegraphics*[width=1.27in, height=1.50in]{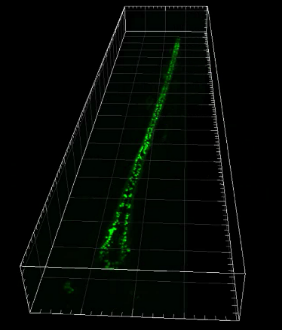}

\vspace{\baselineskip}

\noindent \begin{small} \textbf{Movie\_S8} Legend. Fast growing ipsilateral axon. On the left (contralateral), an ipsilateral axon is seen extending fast at about 120 \textgreek{m}m/h rostral towards caudal at the level of the ventral floor plate along the entire length of the shown 1 dpf rostral spinal cord (viewed at an angle from the top, dorsal side; first visible at t=16 min, bottom left). Length 2 hours in real time, the bar is 30 \textgreek{m}m:\end{small} 

\noindent \includegraphics*[width=2.10in, height=1.30in]{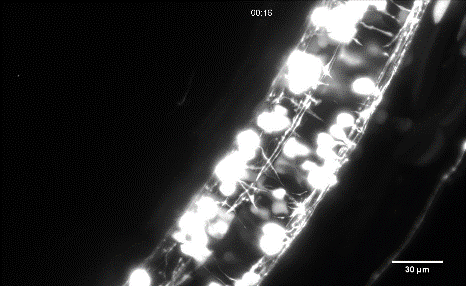}

\vspace{\baselineskip}

\noindent \begin{small} \textbf{Movie\_S9} Legend. Commissural axons avoid the midline in the absence of Robo3, ipsilateral axons are not affected. In a Robo3 morpholino treated zebrafish, a commissural axon is growing toward, next to the asterisk (*), but not crossing the midline (1.5 dpf, caudal half of the zebrafish). Note that the ipsilaterally growing axons are not affected; the strong intensity objects in the middle of the image are cell bodies. Length 4 hours 24 minutes in real time, the bar is 10 \textgreek{m}m: \end{small} 

\noindent \includegraphics*[width=1.55in, height=1.30in]{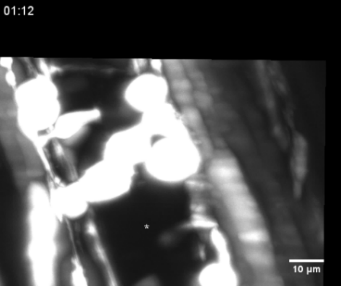}

\vspace{\baselineskip} 

\noindent \begin{small} \textbf{Figure\_S1} Legend: Robo3 Morpholino knock-down and single cell analysis. \textbf{a, b}) Commissural floor plate crossing is dependent on Robo3\textbf{${}^{1}$}: (\textit{a}) Top, residual commissurals observed remaining following the \textit{Robo3} \textit{morpholino} treatment (at 3 dpf; see also Movie S9), arrows mark examples, bar 15 \textgreek{m}m; bottom, commissurals in a \textit{Control} \textit{morpholino} fish (at 5 dpf), arrows mark examples, bar 15 \textgreek{m}m. (\textit{b}) Quantification of the experiment in (a), showing that the number of commissural axons observed to cross the floor plate per \textgreek{m}m spinal cord length following morpholino knock-down of Robo3\textbf{${}^{1}$} is significantly reduced (\textit{Robo3}) (0.0549\underbar{+}0.0224; Morpholino at a total of 4-8 ng/embryo, either only Robo3var2 (the mouse Rig-1 orthologue) or a 1:1 mix of Robo3var2+Robo3var1) compared with the \textit{Control} morpholino (0.1427\underbar{+}0.032; at 8 ng/embryo) (t-test, p$\mathrm{<}$0.01); the accumulated spinal cord length counted for these estimates was 2383 \textgreek{m}m for the Robo3 treatment (n=4 different zebrafish), and 833 \textgreek{m}m for the control (n=3). \textbf{c}) From single-cell FlowSight analysis of dissociation of the spinal cord into single cells, as described in materials and methods, showing examples of the obtained single GFP-positive cell bodies; \textit{Ch01} is brightfield, \textit{Ch02} is GFP at 488 nm. \end{small}

\noindent \includegraphics*[width=6.21in, height=3.80in, trim=0.00in 0.36in 0.00in 0.26in]{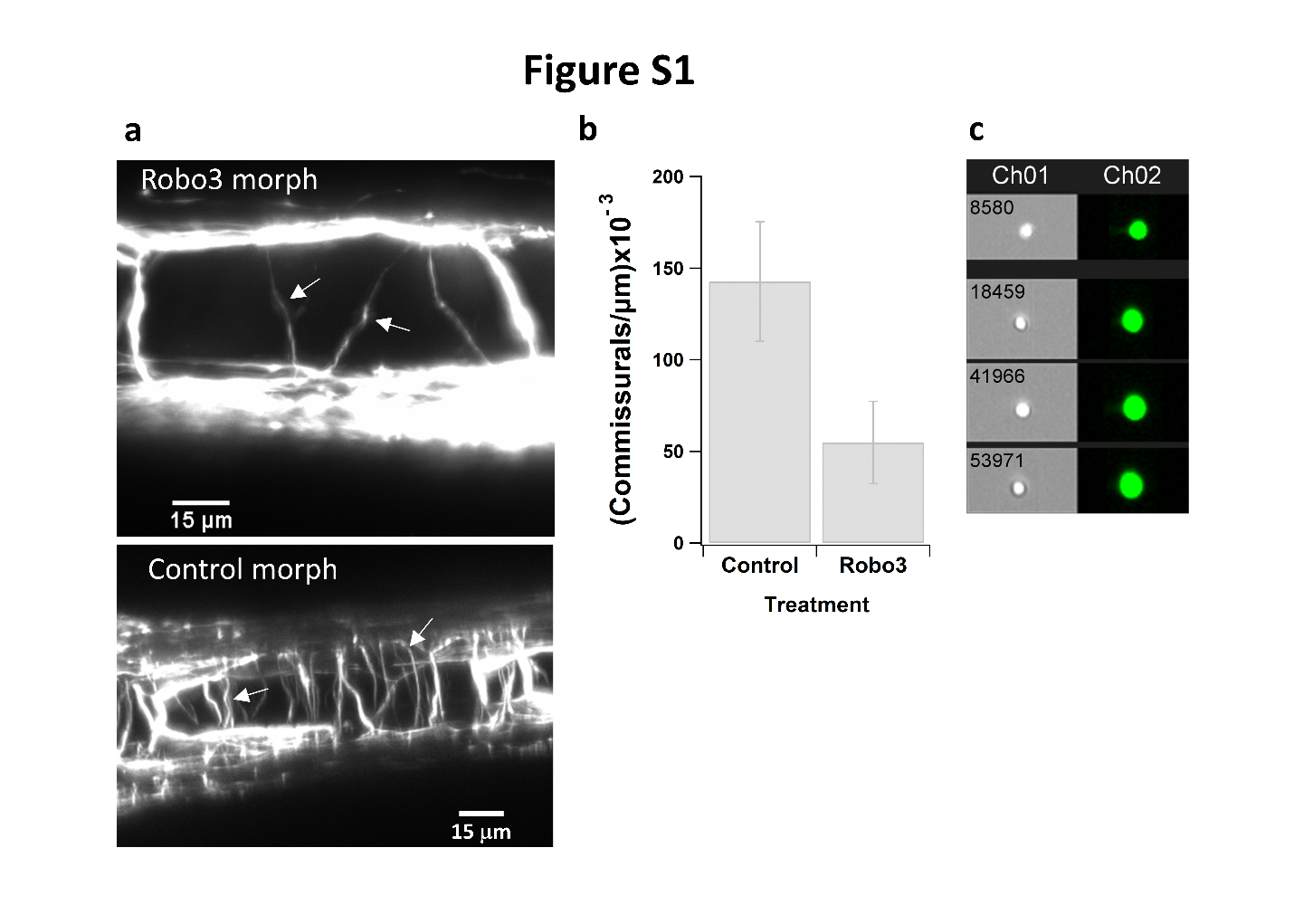}{\centering} 

\noindent \textbf{${}^{1}$}Robo3 has been portrayed as a major midline guidance component. Permitting approach of commissural axons toward the midline and preventing re-crossing once the midline crossed, by regulating somehow the Robo1 and Robo2 receptor activities (Brose et al., 1999; Chen et al., 2008; Friocourt and Chedotal, 2017; Hutson and Chien, 2002; Jaworski et al., 2010; Jaworski et al., 2015; Murray and Whitington, 1999; Wang et al., 1999; Zelina et al., 2014).

\vspace{\baselineskip}

\noindent \textbf{Additional References Supporting Information}

\begin{hangparas}{1.27cm}{1}

Chen, Z., Gore, B. B., Long, H., Ma, L. and Tessier-Lavigne, M. (2008) Alternative Splicing of the Robo3 Axon Guidance Receptor Governs the Midline Switch from Attraction to Repulsion. Neuron 58, 325-332.\\http://dx.doi.org/10.1016/j.neuron.2008.02.016

Friocourt, F. and Chedotal, A. (2017) The Robo3 Receptor, a Key Player in the Development, Evolution, and Function of Commissural Systems. Developmental Neurobiology 77, 876-890. http://dx.doi.org/10.1002/dneu.22478

Jaworski, A., Long, H. and Tessier-Lavigne, M. (2010) Collaborative and Specialized Functions of Robo1 and Robo2 in Spinal Commissural Axon Guidance. The Journal of Neuroscience 30, 9445-9453. http://dx.doi.org/10.1523/jneurosci.6290-09.2010

Wang, K. H., Brose, K., Arnott, D., Kidd, T., Goodman, C. S., Henzel, W. and Tessier-Lavigne, M. (1999) Biochemical Purification of a Mammalian Slit Protein as a Positive Regulator of Sensory Axon Elongation and Branching. Cell 96, 771-784. http://dx.doi.org/10.1016/s0092-8674(00)80588-7

\end{hangparas}

\end{document}